\documentclass[aps,showpacs,floatfix]{revtex4}
\usepackage{amsmath}
\usepackage{graphicx}
\begin{document}


\title{Cluster form factor calculation in the {\it ab initio} no-core shell model
}
\author{Petr Navr\'atil}
\affiliation{Lawrence Livermore National Laboratory, L-414, P.O. Box 808, 
Livermore, CA  94551}

\begin{abstract}
We derive expressions for cluster overlap integrals or channel cluster form factors
for {\it ab initio} no-core shell model (NCSM) wave functions. These are used to obtain 
the spectroscopic factors and can serve as a starting point for the description 
of low-energy nuclear reactions. We consider the composite
system and the target nucleus to be described in the Slater determinant (SD) harmonic
oscillator (HO) basis while the projectile eigenstate to be expanded in the Jacobi coordinate
HO basis. This is the most practical case. The spurious center of mass components
present in the SD bases are removed exactly. The calculated cluster overlap integrals are translationally
invariant. As an illustration, we present results of cluster form factor calculations
for $\langle ^5$He$| ^4$He+n$\rangle$, $\langle ^5$He$| ^3$H+d$\rangle$, $\langle ^6$Li$| ^4$He+d$\rangle$,
$\langle ^6$Be$| ^3$He$+^3$He$\rangle$, $\langle ^7$Li$| ^4$He$+^3$H$\rangle$, 
$\langle ^7$Li$| ^6$Li+n$\rangle$,
$\langle ^8$Be$| ^6$Li+d$\rangle$, $\langle ^8$Be$| ^7$Li+p$\rangle$, $\langle ^9$Li$| ^8$Li+n$\rangle$ 
and $\langle ^{13}$C$| ^{12}$C+n$\rangle$, with all the nuclei described by multi-$\hbar\Omega$ NCSM 
wave functions.
\end{abstract}
\maketitle

\section{Introduction}\label{intro}

There has been a significant progress in the {\it ab initio} approaches
to the structure of light nuclei. Starting from the realistic two- 
and three-nucleon interactions the methods like the Green's Function Monte Carlo (GFMC) \cite{GFMC}
or the {\it ab initio} no-core shell model (NCSM) \cite{C12_NCSM}
can predict the low-lying levels in $p$-shell nuclei. It is a challenging task to extend 
the {\it ab initio} methods to describe nuclear reactions. This is in particular true for low-energy
reactions where detailed knowledge of nuclear structure is important. The first capture reaction 
calculations using the GFMC (or rather variational Monte Carlo) wave functions were performed
\cite{Nollett}. Concerning the NCSM, in order to make the first steps in this direction one needs
to understand the cluster structure of the eigenstates, i.e. to calculate the channel cluster
form factors. Those can then, e.g., be integrated to obtain the spectroscopic factors. At the same time,
starting from the channel cluster form factors one can attempt to set up an approach in the spirit
of the resonating group method (RGM) \cite{RGM} to calculate radial wave functions 
describing the relative motion of the binary clusters and then obtain the cross sections. 
This paper addresses the issue of channel cluster form factor (or cluster overlap integral 
or reduced width amplitude for two-body decay) calculations in the NCSM.

The principal foundation of the {\it ab initio} NCSM  approach is the use of effective 
interactions appropriate for the large but finite basis spaces employed in the
calculations. These effective interactions are derived from the underlying 
realistic inter-nucleon potentials through a unitary transformation in a way 
that guarantees convergence to the exact solution as the basis size increases.
For the basis, one uses antisymmetrized $A$-nucleon harmonic-oscillator (HO) states
that span the complete $N_{\rm max}\hbar\Omega$ space. 
A disadvantage of the HO basis is its unphysical asymptotic behavior, a problem that must be 
dealt with by using a large basis expansion and/or a renormalization. 
On the other hand, the nuclear system is translationally invariant and, in particular
in the case of light nuclei, it is important to preserve this symmetry. The HO basis is the
only basis that allows a switch from Jacobi coordinates to single particle Cartesian coordinates
without violating the translational invariance. Consequently, one may choose the coordinates
according to whatever is more efficient for the problem at hand. In practice, it turns out
that the $A=3$ system is the easiest solved in the Jacobi basis, the $A=4$ system can be solved
either way with the same efficiency when only two-body interaction is utilized, but the Jacobi basis
is more efficient when the three-body interaction is included. For systems with $A>4$, it is 
by far more advantageous to use the Cartesian coordinates and the Slater determinant (SD) 
basis and employ the powerful shell model codes like Antoine \cite{Antoine} that 
rely on the second quantization techniques.
While the NCSM eigenenergies are independent on the choice of coordinates, the eigenfunctions
obtained in the Cartesian coordinate SD basis include a $0\hbar\Omega$ spurious center of mass
(CM) component. 

Our goal is to calculate the channel cluster form factors regardless of the choice 
of coordinates.
Obviously, the most desired case is the one corresponding to the most efficient choice,
i.e., the projectile, that is the lighter nucleus of the binary system, 
consisting of $a\leq 4$ nucleons described by a Jacobi coordinate
wave function, while the $A$-nucleon composite system and the $A-a$ nucleon target, that is
the heavier nucleus of the binary system, described
by wave functions expanded in the SD basis. To obtain the physical, translationally invariant
cluster form factors we must remove completely the spurious CM components.

The ways how to remove these components and obtain physical matrix elements 
of different operators were investigated in the past 
\cite{Tassie,Dieperink,Forest,Philpott,Halderson,Millener1,Millener2}.
At the same time, the single-nucleon as well as cluster overlap integral and/or spectroscopic factor
calculations were investigated in many papers, see e.g. Refs. 
\cite{Smirnov61,Glend65,CK_spec,Rotter68,Kurath73,Ichimura73,KM_3N_spec,Smirnov77,Timof03,Arai04}. 
In many cases, however, the basis space was limited to a single major HO shell.
In the NCSM, the basis space spans several major shells. 
In general, it is necessary to re-visit and adapt the techniques of the channel cluster form factor 
and spectroscopic factor calculation and the spurious 
center-of-mass motion removal to make them applicable for the NCSM.
In an earlier investigation, we addressed the spurious 
center-of-mass motion problem for the density operator \cite{trinv}.
In this paper, we focus on the calculation of the channel cluster form factors.

In Section \ref{sec_calc}, we present the derivation and the algebraic formulas for calculating
the channel cluster form factors from the NCSM wave functions for projectiles consisting of
up to three nucleons. In Section \ref{sec_appl}, applications to several light nuclei systems
are discussed. The conclusions are drawn in Section \ref{sec_concl}. In the Appendix \ref{appB}, 
we give the algebraic cluster form factor expression for the four-nucleon projectile.

\section{Cluster form factor and spectroscopic factor calculation}\label{sec_calc}

In this section we derive expressions for the channel cluster form factors for
a composite system of $A$ nucleons, a projectile of $a$ nucleons and a target of $A-a$ nucleons.
All the nuclei are assumed to be described by eigenstates of the NCSM effective Hamiltonians
expanded in the HO basis with identical HO frequency and the same (for the eigenstates of the 
same parity) or differing by one unit of the HO excitation (for the eigenstates of opposite parity) 
definitions of the model space. We limit ourselves to $a\leq 4$ projectiles. In such a case, 
the projectiles can be efficiently described by a Jacobi-coordinate HO wave functions. The target
and the composite system is assumed to be described by Slater determinant single-particle HO basis
wave functions which is in general more efficient for $A>4$. In this section we present results
for $a=1,2,3$. The cluster overlap integral for the $a=4$ projectile is given in Appendix \ref{appB}.
The NCSM effective interaction theory is not repeated in this paper. It can be found in Ref.
\cite{C12_NCSM} for the case of two-nucleon interactions and in Ref. \cite{NCSM_v3b}
for the case of two- plus three-nucleon interactions.

\subsection{Coordinate and HO wave function transformations}\label{def}

We follow the notation of Ref. \cite{Jacobi_NCSM}.
We consider nucleons with the mass
$m$ neglecting the difference between the proton and the neutron mass.
For the purpose of the present paper we use the following set
of Jacobi coordinates: 
\begin{subequations}\label{jacobiAmaa}
\begin{eqnarray}
\vec{\xi}_0 &=& \sqrt{\frac{1}{A}}\left[\vec{r}_1+\vec{r}_2
                                   +\ldots +\vec{r}_A\right]
\; , \\
\vec{\xi}_1 &=& \sqrt{\frac{1}{2}}\left[\vec{r}_1-\vec{r}_2
                                                     \right]
\; , \\
\vec{\xi}_2 &=& \sqrt{\frac{2}{3}}\left[\frac{1}{2}
                 \left(\vec{r}_1+\vec{r}_2\right)
                                   -\vec{r}_3\right]
\; , \\
&\ldots & \nonumber
\\
\vec{\xi}_{A-a-1} &=& \sqrt{\frac{A-a-1}{A-a}}\left[\frac{1}{A-a-1}
      \left(\vec{r}_1+\vec{r}_2 + \ldots+ \vec{r}_{A-a-1}\right)
                                   -\vec{r}_{A-a}\right]
\; , \\
\vec{\eta}_{A-a} &=& \sqrt{\frac{(A-a)a}{A}}\left[\frac{1}{A-a}
      \left(\vec{r}_1+\vec{r}_2 + \ldots+ \vec{r}_{A-a}\right)
      -\frac{1}{a}\left(\vec{r}_{A-a+1}+\ldots+\vec{r}_{A}\right)\right]
\; , \\
&\ldots &\nonumber
\\
\vec{\vartheta}_{A-2} &=& \sqrt{\frac{2}{3}}\left[\frac{1}{2}
                 \left(\vec{r}_{A-1}+\vec{r}_A\right)
                                   -\vec{r}_{A-2}\right]
\; , \\
\vec{\vartheta}_{A-1} &=& \sqrt{\frac{1}{2}}\left[\vec{r}_{A-1}-\vec{r}_A
                                                     \right]
\; .
\end{eqnarray}
\end{subequations}
Here, $\vec{\xi}_0$ is proportional to the center of mass of the
$A$-nucleon system: $\vec{R}=\sqrt{\frac{1}{A}}\vec{\xi}_0$. 
On the other hand, $\vec{\xi}_\rho$ is proportional
to the relative position of the $\rho+1$-st nucleon and the
center of mass of the $\rho$ nucleons. The $\vec{\eta}_{A-a}$ coordinate
is proportional to the relative position between the center of masses of the two
interacting clusters, i.e. the $A-a$ nucleon target and the $a$-nucleon projectile.
The $\vec{\vartheta}$ coordinates appear only for $a>1$.
Let us rewrite the Eq. (\ref{jacobiAmaa}e) and (\ref{jacobiAmaa}a) as
\begin{subequations}\label{jacobi_tr}
\begin{eqnarray}
\vec{\eta}_{A-a} &=& \sqrt{\frac{a}{A}} \vec{R}^{A-a}_{\rm CM}
                    -\sqrt{\frac{A-a}{A}}\vec{R}^{a}_{\rm CM}
\; , \\
\vec{\xi}_0 &=& \sqrt{\frac{A-a}{A}} \vec{R}^{A-a}_{\rm CM}
                    +\sqrt{\frac{a}{A}}\vec{R}^{a}_{\rm CM}
\; ,
\end{eqnarray}
\end{subequations}
where 
$\vec{R}^{A-a}_{\rm CM}=\sqrt{\frac{1}{A-a}}\left[\vec{r}_1+\vec{r}_2+\ldots +\vec{r}_{A-a}\right]$
and $\vec{R}^{a}_{\rm CM}=\sqrt{\frac{1}{a}}\left[\vec{r}_{A-a+1}+\ldots +\vec{r}_{A}\right]$
Following, e.g. Ref. \cite{Tr72}, the HO wave functions depending 
on the coordinates (\ref{jacobi_tr}) transform as
\begin{eqnarray}\label{ho_tr}
&&\sum_{M m} (L M l m|Q q) \varphi_{N L M}(\vec{R}^{A-a}_{\rm CM}) 
\varphi_{n l m}(\vec{R}^{a}_{\rm CM}) = 
\nonumber \\
&&\sum_{n' l' m' N' L' M'} \langle n'l' N'L' Q|N L n l Q\rangle_{\frac{a}{A-a}}
(l' m' L' M'|Q q) \varphi_{n'l'm'}(\vec{\eta}_{A-a}) \varphi_{N'L'M'}(\vec{\xi}_0)
\; ,
\end{eqnarray}
where $\langle n'l' N'L' Q|N L n l Q\rangle_{\frac{a}{A-a}}$ 
is the general HO bracket for two particles with mass ratio $\frac{a}{A-a}$.

\subsection{Composite and asymptotic wave functions and the channel cluster form factor}\label{states}

We consider the $A$-nucleon composite state eigenfunction 
\begin{equation}\label{comp_state}
\langle \vec{\xi}_1 \ldots \vec{\eta}_{A-a} \ldots \vec{\vartheta}_{A-1} 
\sigma_1 \ldots \sigma_A \tau_1 \ldots \tau_A | A \lambda J M T M_T\rangle \;
\end{equation}
with the $\sigma$ and $\tau$ the spin and isospin coordinates, respectively. The $J$ and $T$ 
is the total angular momentum and the total isospin, respectively, and $M, M_T$ their 
third components.The $\lambda$
stands for the additional quantum numbers needed to characterize the eigenstate.
The $\vec{\vartheta}$ coordinates appear only for $a>1$.

Projectile-target wave function with the radial wave function describing the relative
motion of the two nuclei replaced by the Dirac delta function can be written as
%
%
\begin{eqnarray}\label{proj-targ_state_delta}
\langle\vec{\xi}_1 \ldots \vec{\xi}_{A-a-1} \eta^\prime_{A-a} &\hat{\eta}_{A-a}& 
\vec{\vartheta}_{A-a+1} \ldots \vec{\vartheta}_{A-1} 
\sigma_1 \ldots \sigma_A \tau_1 \ldots \tau_A 
|\Phi_{\alpha I_1 T_1, \beta I_2 T_2;s l}^{(A-a,a)J M T M_T};\delta_{\eta_{A-a}}\rangle  
\nonumber \\ 
&=&\sum (I_1 M_1 I_2 M_2 | s m_s) (s m_s l m_l | J M) (T_1 M_{T_1} T_2 M_{T_2} | T M_T) 
\frac{\delta(\eta_{A-a}-\eta^\prime_{A-a})}{\eta_{A-a}\eta^\prime_{A-a}} 
Y_{l m_l}(\hat{\eta}_{A-a}) 
\nonumber \\ 
&\times&
\langle \vec{\xi}_1 \ldots \vec{\xi}_{A-a-1} \sigma_1 \ldots \sigma_{A-a} 
\tau_1 \ldots \tau_{A-a} | A-a \alpha I_1 M_1 T_1 M_{T_1}\rangle
\nonumber \\ 
&\times& \langle \vec{\vartheta}_{A-a+1} \ldots \vec{\vartheta}_{A-1} \sigma_{A-a+1} \ldots \sigma_{A} 
\tau_{A-a+1} \ldots \tau_{A} | a \beta I_2 M_2 T_2 M_{T_2}\rangle
\; ,
\end{eqnarray}
where $\langle \vec{\xi}_1 \ldots \vec{\xi}_{A-a-1} \sigma \tau | A-a \alpha I_1 M_1 T_1 M_{T_1}\rangle $ 
and
$\langle \vec{\vartheta}_{A-a+1} \ldots \vec{\vartheta}_{A-1}\sigma\tau | a\beta I_2 M_2 T_2 M_{T_2}\rangle $ 
are the target and the projectile eigenstates, respectively. 
The $s$ is the channel spin and the $l$ is the channel relative orbital angular momentum.
For our convenience, we also define a projectile-target wave functions with a HO radial
wave function describing relative motion of the two nuclei, i.e.
\begin{eqnarray}\label{proj-targ_state_HO}
\langle\vec{\xi}_1 \ldots \vec{\xi}_{A-a-1} &\vec{\eta}_{A-a}& 
\vec{\vartheta}_{A-a+1} \ldots \vec{\vartheta}_{A-1} 
\sigma_1 \ldots \sigma_A \tau_1 \ldots \tau_A 
|\Phi_{\alpha I_1 T_1, \beta I_2 T_2;s l}^{(A-a,a)J M T M_T};nl\rangle  
\nonumber \\ 
&=&\sum (I_1 M_1 I_2 M_2 | s m_s) (s m_s l m_l | J M) (T_1 M_{T_1} T_2 M_{T_2} | T M_T) 
R_{nl}(\eta_{A-a})
Y_{l m_l}(\hat{\eta}_{A-a}) 
\nonumber \\ 
&\times&
\langle \vec{\xi}_1 \ldots \vec{\xi}_{A-a-1} \sigma_1 \ldots \sigma_{A-a} 
\tau_1 \ldots \tau_{A-a} | A-a \alpha I_1 M_1 T_1 M_{T_1}\rangle
\nonumber \\ 
&\times& \langle \vec{\vartheta}_{A-a+1} \ldots \vec{\vartheta}_{A-1} \sigma_{A-a+1} \ldots \sigma_{A} 
\tau_{A-a+1} \ldots \tau_{A} | a \beta I_2 M_2 T_2 M_{T_2}\rangle
\; .
\end{eqnarray}
The $R_{nl}(r)$ in Eq. (\ref{proj-targ_state_HO}) is the radial HO 
wave function with the oscillator length parameter $b=b_0=\sqrt{\frac{\hbar}{m\Omega}}$, 
where $m$ is the nucleon mass. Due to our use of the coordinate 
transformations (\ref{jacobiAmaa}) the oscillator length parameter is the same 
for all coordinates, i.e. $b_0$.
In Eqs. (\ref{proj-targ_state_delta},\ref{proj-targ_state_HO}), 
the coordinates $\vec{\vartheta}$ appear only for $a>1$.

The channel cluster form factor can then be defined as
\begin{eqnarray}\label{cluster_form_factor}
u^{A\lambda JT}_{A-a \alpha I_1 T_1, a \beta I_2 T_2; s l}(\eta_{A-a})&=&
\langle A \lambda J T|{\cal A}\Phi_{\alpha I_1 T_1, \beta I_2 T_2;s l}^{(A-a,a)J T};
\delta_{\eta_{A-a}}\rangle
= \sum_n R_{nl}(\eta_{A-a}) 
\langle A \lambda J T|{\cal A}\Phi_{\alpha I_1 T_1, \beta I_2 T_2;s l}^{(A-a,a)J T};
nl \rangle
\nonumber \\
&=& \sqrt{\frac{A!}{(A-a)!a!}} \sum_n R_{nl}(\eta_{A-a}) 
\langle A \lambda J T|\Phi_{\alpha I_1 T_1, \beta I_2 T_2;s l}^{(A-a,a)J T};
nl \rangle
\; ,
\end{eqnarray}
with ${\cal A}$ the antisymmetrizer. As stated above, we assume identical HO frequency for all
eigenstates and identical (or differing by a single HO excitation in the case of opposite 
parity states) definitions of the model space.

The spectroscopic factor is obtained by integrating the square of the cluster form
factor. In particular, we have
\begin{eqnarray}\label{spec_fac}
S^{A\lambda JT}_{A-a \alpha I_1 T_1, a \beta I_2 T_2; s l}&=&
\int d\eta_{A-a} \eta_{A-a}^2 
|u^{A\lambda JT}_{A-a \alpha I_1 T_1, a \beta I_2 T_2; s l}(\eta_{A-a})|^2
\nonumber \\
&=& \frac{A!}{(A-a)!a!} \sum_n 
|\langle A \lambda J T|\Phi_{\alpha I_1 T_1, \beta I_2 T_2;s l}^{(A-a,a)J T};
nl \rangle|^2
\; .
\end{eqnarray}
As in this paper all the eigenstates are assumed to be expanded in a large but finite HO basis,
we can set the integration limit to infinity in Eq. (\ref{spec_fac}).

It turns out that obtaining the eigenstates using the Jacobi coordinates becomes increasingly difficult
with the number of nucleons $A$ mostly due to the complicated antisymmetrization. 
As stated in the Introduction,
for $A>4$ it is by far more efficient to use the SD basis. Consequently, it is desirable to
to express the overlap (\ref{cluster_form_factor}) using the eigenstates obtained in the SD basis.

The relationship between the Jacobi coordinate and the SD eigenstates is 
\begin{equation}\label{state_relation}
\langle \vec{r}_1 \ldots \vec{r}_A \sigma_1 \ldots \sigma_A \tau_1 \ldots \tau_A 
| A \lambda J M T M_T\rangle_{\rm SD} = 
\langle \vec{\xi}_1 \ldots \vec{\eta}_{A-a} \ldots \vec{\vartheta}_{A-1} 
\sigma_1 \ldots \sigma_A \tau_1 \ldots \tau_A 
| A \lambda J M T M_T\rangle \varphi_{000}(\vec{\xi}_0) \; ,
\end{equation}
for the composite system and similarly for the $A-a$ nucleon target.
The subscript SD refers to the fact that this state was obtained 
in the Slater determinant basis, i.e. by using
a shell model code, and, consequently, contains the spurious CM component. 

To arrive at the desired expression, we investigate an analogues overlap to 
(\ref{cluster_form_factor}) using the SD eigenstates. We consider
corresponding SD eigenstates to (\ref{comp_state}) and (\ref{proj-targ_state_HO}), i.e., 
\begin{equation}\label{SD_overlap}
_{\rm SD}\langle A\lambda J T| {\cal A} 
\Phi_{\alpha I_1 T_1, \beta I_2 T_2;s l}^{(A-a,a)J T};nl\rangle_{\rm SD} \; ,  
\end{equation}
where all the composite and the target eigenstate Jacobi coordinates are replaced 
by the Cartesian coordinates. The projectile eigenstate is kept unchanged with the Jacobi 
coordinates. Further, the $\vec{\eta}_{A-a}$ is replaced by $\vec{R}_{\rm CM}^a$.
Explicitly, we have for the SD analog of the state (\ref{proj-targ_state_HO})
\begin{eqnarray}\label{SD_proj-targ_state_HO}
\langle\vec{r}_1 \ldots \vec{r}_{A-a} &\vec{R}^{a}_{\rm CM}& 
\vec{\vartheta}_{A-a+1} \ldots \vec{\vartheta}_{A-1} 
\sigma_1 \ldots \sigma_A \tau_1 \ldots \tau_A 
|\Phi_{\alpha I_1 T_1, \beta I_2 T_2;s l}^{(A-a,a)J M T M_T};nl\rangle_{\rm SD} \;
\nonumber \\ 
&=&\sum (I_1 M_1 I_2 M_2 | s m_s) (s m_s l m_l | J M) (T_1 M_{T_1} T_2 M_{T_2} | T M_T) 
R_{nl}(R^{a}_{\rm CM})
Y_{l m_l}(\hat{R}^{a}_{\rm CM}) 
\nonumber \\ 
&\times&
\langle \vec{r}_1 \ldots \vec{r}_{A-a} \sigma_1 \ldots \sigma_{A-a} 
\tau_1 \ldots \tau_{A-a} | A-a \alpha I_1 M_1 T_1 M_{T_1}\rangle_{\rm SD} \;
\nonumber \\ 
&\times& \langle \vec{\vartheta}_{A-a+1} \ldots \vec{\vartheta}_{A-1} \sigma_{A-a+1} 
\ldots \sigma_{A} \tau_{A-a+1} \ldots \tau_{A} | a \beta I_2 M_2 T_2 M_{T_2}\rangle
\; .
\end{eqnarray}
We now proceed in two steps. First, using the relation (\ref{state_relation}) for both
the composite and the target eigenstate and the transformation (\ref{ho_tr}) we obtain
\begin{equation}\label{SD_Jacobi_overlap}
_{\rm SD}\langle A\lambda J T| {\cal A} 
\Phi_{\alpha I_1 T_1, \beta I_2 T_2;s l}^{(A-a,a)J T};nl\rangle_{\rm SD} \;
= \langle nl00l|00nll\rangle_{\frac{a}{A-a}} \;
\langle A\lambda J T| {\cal A} 
\Phi_{\alpha I_1 T_1, \beta I_2 T_2;s l}^{(A-a,a)J T};nl\rangle
\; ,  
\end{equation}
with a general HO bracket due to the CM motion, which value is simply given by
\begin{equation}\label{cm_ho_br}
\langle nl00l|00nll\rangle_{\frac{a}{A-a}} = (-1)^l 
\left(\frac{A-a}{A}\right)^{\frac{2n+l}{2}}
\; .
\end{equation}
The relation (\ref{SD_Jacobi_overlap}) has been derived in the past, see e.g. Refs. \cite{Smirnov61},
\cite{Rotter68}, \cite{Smirnov77}. 
Second, we relate the overlap (\ref{SD_overlap}) to a linear combination of matrix elements 
of $a$ creation operators between the target and the composite eigenstates
$_{\rm SD}\langle A\lambda J T|a^\dagger_{n_1l_1j_1}\ldots a^\dagger_{n_al_aj_a}
| A-a \alpha I_1 T_1\rangle_{\rm SD}$. The subscripts $n_1 l_1 j_1$ refer to the single-particle
state quantum numbers $n_1 (l_1\frac{1}{2}) j_1 m_1 \frac{1}{2} m_{t_1}$ etc. 
Such matrix elements are easily calculated by shell model codes.
To obtain the channel cluster form factor we use the second equality 
in Eq. (\ref{cluster_form_factor}).

\subsection{Single-nucleon projectile}\label{a=1}

In the case of a single-nucleon projectile, the asymptotic state (\ref{proj-targ_state_delta})
simplifies as no $\vec{\vartheta}$ coordinates are present. The projectile wave function
has just spin and isospin components with $I_2=\frac{1}{2}$ and $T_2=\frac{1}{2}$, respectively. 
It is straightforward 
to calculate the overlap of the states (\ref{comp_state}) and (\ref{proj-targ_state_delta}).
The result is given by
\begin{eqnarray}\label{single-nucleon_Jacobi}
\langle A \lambda J T|&{\cal A}& \Phi_{\alpha I_1 T_1, \textstyle{\frac{1}{2}}
\textstyle{\frac{1}{2}}; s l}^{(A-1,1) JT};\delta_{\eta_{A-1}}\rangle 
= \sqrt{A} \sum R_{nl}(\eta_{A-1})
\hat{s}\hat{j}(-1)^{I_1+J+j}
\left\{ \begin{array}{ccc} I_1 & \textstyle{\frac{1}{2}} & s \\
  l  & J & j   
\end{array}\right\}
\nonumber \\
&\times&
\langle A \lambda J T|(N_{A-1} i_{A-1} I_1 T_1; n l j \textstyle{\frac{1}{2}}) J T\rangle
\langle N_{A-1} i_{A-1} I_1 T_1 | A-1 \alpha I_1 T_1\rangle
\; ,
\end{eqnarray}
with $\hat{s}=\sqrt{2s+1}$.
The composite eigenstate is expanded in a basis with lower degree of antisymmetry
using the coefficients of fractional parentage \cite{Jacobi_NCSM} 
\begin{equation}\label{A-1_Jacobi_state}
\langle (N_{A-1} i_{A-1} I_1 T_1; n l j \textstyle{\frac{1}{2}}) J T | A \lambda J T\rangle
=\sum_{N i} \langle N_{A-1} i_{A-1} I_1 T_1; n l j || N i J T \rangle
\langle N i J T | A \lambda J T\rangle \; ,
\end{equation}
with $N=N_{\rm A-1} + 2n + l$ the total number of HO excitations for the $A$ nucleons
and $i,i_{\rm A-1}$ the additional quantum numbers that characterize the $A$- and $A-1$-nucleon
antisymmetrized basis states, respectively. 

To obtain the overlap integral matrix element starting from SD composite and target eigenstates,
we make use of (\ref{SD_Jacobi_overlap}) with $a=1$ and perform the above discussed second step.
That is quite straightforward for the $a=1$ case and we easily arrive at the final expression: 
\begin{eqnarray}\label{single-nucleon}
\langle A \lambda J T|&{\cal A}& \Phi_{\alpha I_1 T_1, \textstyle{\frac{1}{2}}
\textstyle{\frac{1}{2}}; s l}^{(A-1,1) JT};\delta_{\eta_{A-1}}\rangle 
= \sum_n R_{nl}(\eta_{A-1}) 
\frac{1}{\langle nl00l|00nll\rangle_{\frac{1}{A-1}}} \frac{1}{\hat{J}\hat{T}} 
\sum_j \hat{s}\hat{j}(-1)^{I_1+J+j}
\left\{ \begin{array}{ccc} I_1 & \textstyle{\frac{1}{2}} & s \\
  l  & J & j   
\end{array}\right\}
\nonumber \\
&\times&
\; _{\rm SD}\langle A\lambda JT|||a^\dagger_{nlj}|||A-1\alpha I_1 T_1\rangle_{\rm SD} \;
\; .
\end{eqnarray}
Using Eq. (\ref{cm_ho_br}), we obtain the familiar CM correction factor $(\frac{A}{A-1})^{\frac{2n+l}{2}}$
\cite{Dieperink,Forest,Philpott,Halderson,Millener1,Millener2,Smirnov61}.

\subsection{Two-nucleon projectile}\label{a=2}

For $a>1$ projectiles we only present the overlap matrix elements for the composite and target
wave functions obtained in the SD basis. For $a=2$, which includes the deuteron projectile,
the derivation is slightly more complicated due to additional re-couplings and explicit presence 
of the $a=2$ relative coordinate wave function expanded in the HO basis. The final expression
reads:
\begin{eqnarray}\label{two-nucleon}
\langle A \lambda JT |&{\cal A}& \Phi_{\alpha I_1 T_1,\beta I_2 T_2;s l}^{(A-2,2) JT};
\delta_{\eta_{A-2}}\rangle
= \sum_n R_{nl}(\eta_{A-2}) \frac{1}{\sqrt{2}} 
\frac{1}{\langle nl00l|00nll\rangle_{\frac{2}{A-2}}} \frac{1}{\hat{J}\hat{T}} 
\sum \langle n_2 l_2 s_2 I_2 T_2|a=2 \beta I_2 T_2 \rangle
\nonumber \\
&\times& 
\hat{s}\hat{s}_2\hat{I}_2\hat{j}_a\hat{j}_b \hat{I}_{ab}\hat{L}_{ab}^2 (-1)^{I_1+J+l+l_2+T_2}
\left\{ \begin{array}{ccc} I_1 & I_2 & s \\
  l  & J & I_{ab}   
\end{array}\right\}
\left\{ \begin{array}{ccc} l & L_{ab} & l_2 \\
  s_2  & I_2 & I_{ab}   
\end{array}\right\}
\left\{ \begin{array}{ccc} 
     l_a & l_b & L_{ab} \\
  \textstyle{\frac{1}{2}} & \textstyle{\frac{1}{2}} & s_2 \\
     j_a & j_b & I_{ab}
\end{array}\right\}
\nonumber \\
&\times&
\langle n_a l_a n_b l_b L_{ab}|nl n_2 l_2 L_{ab}\rangle_{1}
\; _{\rm SD}\langle A\lambda JT|||(a^\dagger_{n_a l_a j_a} a^\dagger_{n_b l_b j_b})
^{(I_{ab} T_2)}
|||A-2\alpha I_1 T_1\rangle_{\rm SD} \;
\; ,
\end{eqnarray}
with the antisymmetry condition for the two-nucleon channels $(-1)^{l_2+s_2+T_2}=-1$.
The two-nucleon projectile wave function 
$\langle n_2 l_2 s_2 I_2 T_2|a=2 \beta I_2 T_2 \rangle $ is expanded in the 
HO basis depending on $\vec{\vartheta}_{A-1}$. The spin and isospin components of the wave 
function depend on the spin and isopin coordinates $\sigma_{A-1}\sigma_A$, and $\tau_{A-1},\tau_A$,
respectively. 
For the deuteron projectile, $I_2=1, T_2=0, s_2=1$ and $l_2=0\;\; or\;\; 2$. 
Here, in addition to the HO bracket (\ref{cm_ho_br}) due 
to the CM correction, one more HO bracket appears that corresponds
to particles with mass ratio $1$. This is due to the transformation of the HO wave functions
$\varphi_{nlm}(\vec{R}_{\rm CM}^{a=2}) \varphi_{n_2 l_2 m_2}(\vec{\vartheta}_{A-1})$ 
to the single-particle HO wave functions 
$\varphi_{n_a l_a m_a}(\vec{r}_A) \varphi_{n_b l_b m_b}(\vec{r}_{A-1})$.

\subsection{Three-nucleon projectile}\label{a=3}

For $a=3$, which includes the triton or $^3$He projectile,
the derivation is still more complicated due to additional re-couplings and explicit presence 
of the $a=3$ relative coordinate wave function expanded in the HO basis. The final expression
reads:
\begin{eqnarray}\label{three-nucleon}
\langle A \lambda JT |&{\cal A}& \Phi_{\alpha I_1 T_1, \beta I_2 T_2; s l}^{(A-3,3) JT};
\delta_{\eta_{A-3}}\rangle
= \sum_n R_{nl}(\eta_{A-3}) \frac{1}{\sqrt{6}} 
\frac{1}{\langle nl00l|00nll\rangle_{\frac{3}{A-3}}} \frac{1}{\hat{J}\hat{T}} 
\nonumber \\
&\times&
\sum \langle (n_2 l_2 s_2 j_2 t_2;{\cal N}_2 {\cal L}_2 {\cal J}_2 \textstyle{\frac{1}{2}}) 
I_2 T_2|a=3 \beta I_2 T_2 \rangle
\nonumber \\
&\times& 
\hat{s}\hat{I}\hat{s}_2 \hat{j}_2 \hat{I}_2 \hat{\cal J}_2 
\hat{j}_a\hat{j}_b \hat{j}_c \hat{I}_{ab} \hat{\lambda}^2 \hat{L}_{ab}^2 
(-1)^{I_1-I+J+l_c+l+{\cal J}_2+\textstyle{\frac{1}{2}}+l_2+t_2+I_{ab}}
\left\{ \begin{array}{ccc} I_1 & I_2 & s \\
  l  & J & I   
\end{array}\right\}
\nonumber \\
&\times&
\left\{ \begin{array}{ccc} L_2 & L_{ab} & l_2 \\
  s_2  & j_2 & I_{ab}   
\end{array}\right\}
\left\{ \begin{array}{ccc} 
     l_a & l_b & L_{ab} \\
  \textstyle{\frac{1}{2}} & \textstyle{\frac{1}{2}} & s_2 \\
     j_a & j_b & I_{ab}
\end{array}\right\}
\left\{ \begin{array}{cccc} 
     l & \lambda & L_2 & j_2 \\
   {\cal L}_2  & l_c & I_{ab} & I_2 \\
   {\cal J}_2 & \textstyle{\frac{1}{2}} & j_c & I
\end{array}\right\}
\nonumber \\
&\times&
\langle n_a l_a n_b l_b L_{ab}|N_2 L_2 n_2 l_2 L_{ab}\rangle_{1}
\langle n_c l_c N_2 L_2 \lambda|nl {\cal N}_2 {\cal L}_2 \lambda \rangle_{\textstyle{\frac{1}{2}}}
\nonumber \\
&\times&
\; _{\rm SD}\langle A\lambda JT|||((a^\dagger_{n_a l_a j_a} a^\dagger_{n_b l_b j_b})^{(I_{ab} t_2)}
a^\dagger_{n_c l_c j_c})^{(I T_2)}
|||A-3\alpha I_1 T_1\rangle_{\rm SD} 
\; ,
\end{eqnarray}
The three-nucleon eigenstates are expanded in a basis with lower degree of antisymmetry
using the coefficients of fractional parentage \cite{Jacobi_NCSM}
\begin{equation}\label{cfp_3}
\langle (n_2 l_2 s_2 j_2 t_2;{\cal N}_2 {\cal L}_2 {\cal J}_2 \textstyle{\frac{1}{2}}) 
I_2 T_2|a=3 \beta I_2 T_2 \rangle = \sum_{N i} 
\langle n_2 l_2 s_2 j_2 t_2;{\cal N}_2 {\cal L}_2 {\cal J}_2 \textstyle{\frac{1}{2}} 
|| N i I_2 T_2 \rangle \langle N i I_2 T_2 | a=3 \beta I_2 T_2 \rangle 
\; ,
\end{equation}
with $N=2{\cal N}_2+{\cal L}_2 + 2n_2 + l_2$ the total number of HO excitations 
for the three nucleons and $i$ the additional quantum number that characterizes 
the three-nucleon antisymmetrized basis states. The $12-j$ symbol of the first kind 
\cite{Varshalovich} appearing in Eq. ({\ref{three-nucleon}) is defined in Appendix \ref{appA}.
For the triton or $^3$He projectile, $I_2=\frac{1}{2}, T_2=\frac{1}{2}$ and 
$(-1)^{l_2+{\cal L}_2}=1$.
In Eq. (\ref{three-nucleon}), in addition to the HO bracket (\ref{cm_ho_br}) due 
to the CM correction, two general HO brackets appear that correspond
to particles with mass ratios $1$ and $\frac{1}{2}$. 
These are due to the sequence of two transformations of the HO wave functions
$\varphi_{nlm}(\vec{R}_{\rm CM}^{a=3}) \varphi_{n_2 l_2 m_2}(\vec{\vartheta}_{A-1}) 
\varphi_{{\cal N}_2 {\cal L}_2 {\cal M}_2}(\vec{\vartheta}_{A-2})$ 
to the single-particle HO wave functions 
$\varphi_{n_a l_a m_a}(\vec{r}_A) \varphi_{n_b l_b m_b}(\vec{r}_{A-1}) 
\varphi_{n_c l_c m_c}(\vec{r}_{A-2})$.

\section{Applications}\label{sec_appl}

In this section, we present results of cluster form factor and/or spectroscopic factor 
calculations 
for $\langle ^5$He$| ^4$He+n$\rangle$, $\langle ^5$He$| ^3$H+d$\rangle$, 
$\langle ^6$Li$| ^4$He+d$\rangle$,
$\langle ^6$Be$| ^3$He$+^3$He$\rangle$, $\langle ^7$Li$| ^4$He$+^3$H$\rangle$, 
$\langle ^7$Li$| ^6$Li+n$\rangle$,
$\langle ^8$Be$| ^6$Li+d$\rangle$, $\langle ^8$Be$| ^7$Li+p$\rangle$, 
$\langle ^9$Li$| ^8$Li+n$\rangle$ and
$\langle ^{13}$C$| ^{12}$C+n$\rangle$. All calculations are done using the approach described 
in Sect. \ref{sec_calc}. The composite $A$-nucleon system and the target $A-a$ nucleon
system are described by the NCSM wave functions obtained in the m-scheme Slater-determinant 
basis shell model calculation. In particular, we use the many-fermion dynamics (MFD) 
shell model code \cite{MFD} and a specialized transition density code that calculates the 
$\langle a^\dagger\ldots a^\dagger\rangle$ matrix elements employing the wave functions 
obtained by the MFD (or the Antoine \cite{Antoine}) code.
The projectile $a$-nucleon NCSM wave functions for $a=3$ are obtained in the Jacobi-coordinate 
HO basis using the code MANYEFF \cite{Jacobi_NCSM}. For the $a=2$, i.e. the deuteron 
projectile, the relative-coordinate wave function is obtained using the standard NCSM two-body 
effective interaction code, see e.g. Ref. \cite{C12_NCSM}. As a technical point, we note that
in the case of $a=2$ there is no CM HO binding potential contrary the usual NCSM
two-body effective interaction calculation. Consequently, the overlap of the full-space 
$a=2$ wave function with the model space $P$ might not be large. This then could lead 
to numerical difficulties when applying the Lee-Suzuki procedure \cite{LS1,UMOA} 
to obtain the model space $a=2$ wave functions. To address this issue, higher precision 
than the double precision had to be used in the relevant computer code. 

We performed several calculations to test correctness of the formulas presented 
in Sect. \ref{sec_calc} as well as their computer coding. First, we cross checked that the
Eq. (\ref{single-nucleon_Jacobi}) and Eq. (\ref{single-nucleon}) give the same result
for the $\langle ^5$He$| ^4$He+n$\rangle$ system. In the former case, we employed the Jacobi-coordinate
MANYEFF code while in the latter we used the SD basis MFD shell model code together with the
transition density code. Obviously, 
the same effective Hamiltonian was used in both calculations. To test the Eqs. 
(\ref{two-nucleon},\ref{three-nucleon}) we switched the role of the projectile and the target.
For example, for the $\langle ^3$H$| $d+n$\rangle$ system, we can apply Eq. (\ref{single-nucleon})
with the deuteron as the target and the neutron as the projectile or we can apply Eq. 
(\ref{two-nucleon}) with the neutron as the target and the deuteron as the projectile.
In the latter case, the neutron target state is described as a single neutron $N=0$
HO state. Similarly, the Eq. (\ref{three-nucleon}) can be tested considering the
$\langle ^4$He$| ^3$He+n$\rangle$ system described with the neutron as the projectile
in Eq. (\ref{single-nucleon}) or $^3$He as the projectile in Eq. (\ref{three-nucleon}).
Finally, we also performed a test for the $\langle ^5$He$| ^3$H+d$\rangle$ system
by switching the target and projectile and using the Eqs. (\ref{two-nucleon}) and
(\ref{three-nucleon}), respectively. We note that all these tests are non-trivial
as the projectile and the target are described using different coordinates and the
respective wave functions are obtained by different computer codes.
  
In this Section, all the calculated channel cluster form factors are presented as 
a function of the separation $r$ between the CM of the projectile and the CM of the target.
In particular, $r=\sqrt{\frac{A}{(A-a)a}} \eta_{A-a}$ with $\vec{\eta}_{A-a}$ defined in
Eq. (\ref{jacobiAmaa}e). Consequently, we have to use the
reduced mass $\mu=\frac{(A-a)a}{A}m$ in the definition of the HO length parameter, 
$b=\sqrt{\frac{\hbar}{\mu\Omega}}=\sqrt{\frac{A}{(A-a)a}} b_0$. The presented channel
cluster form factors are then related to those defined in Eq. (\ref{cluster_form_factor})
by $u^{A\lambda JT}_{A-a \alpha I_1 T_1, a \beta I_2 T_2; s l}(r)=
\sum_n R_{nl}(r,b) 
\langle A \lambda J T|{\cal A}\Phi_{\alpha I_1 T_1, \beta I_2 T_2;s l}^{(A-a,a)J T};
nl \rangle=
\sum_n (\frac{(A-a)a}{A})^{3/2} R_{nl}(\eta_{A-a},b_0) 
\langle A \lambda J T|{\cal A}\Phi_{\alpha I_1 T_1, \beta I_2 T_2;s l}^{(A-a,a)J T};
nl \rangle $.
The spectroscopic factor (\ref{spec_fac}) is not affected by the choice of the coordinate:
$S^{A\lambda JT}_{A-a \alpha I_1 T_1, a \beta I_2 T_2; s l}=
\int d\eta_{A-a} \eta_{A-a}^2 
|u^{A\lambda JT}_{A-a \alpha I_1 T_1, a \beta I_2 T_2; s l}(\eta_{A-a})|^2
=\int dr r^2 
|u^{A\lambda JT}_{A-a \alpha I_1 T_1, a \beta I_2 T_2; s l}(r)|^2
=\sum_n 
|\langle A \lambda J T|{\cal A} \Phi_{\alpha I_1 T_1, \beta I_2 T_2;s l}^{(A-a,a)J T};
nl \rangle|^2 $. The channel cluster form factors presented in this section are obtained
from Eqs. (\ref{single-nucleon},\ref{two-nucleon},\ref{three-nucleon}) with the radial 
HO wave function $R_{nl}(\eta_{A-a},b_0)$ replaced by $R_{nl}(r,b)$.

\subsection{$\langle ^5$He$| ^4$He+n$\rangle$}\label{he5_he4}

In Figs. \ref{he5_he4_nmax} and \ref{he5_he4_omega}, we present our
$\langle ^5$He$| ^4$He+n$\rangle$ results for the $^5$He $\frac{3}{2}^-$
ground state resonance. The dependence of the channel cluster form factor
on the basis size of the NCSM calculation is shown in Fig. \ref{he5_he4_nmax}
for basis sizes from $N_{\rm max}=4$ $(4\hbar\Omega)$ to 
$N_{\rm max}=12$ $(12\hbar\Omega)$. Here, $N_{\rm max}$ is
the maximal number of the HO excitations above the unperturbed ground state.
The CD-Bonn 2000 NN potential \cite{cdb2k} and the HO frequency of $\hbar\Omega=16$ MeV
were used in the calculations. Clearly, with increasing $N_{\rm max}$, the changes
between successive curves become smaller, a sign of convergence. 

When calculating overlaps involving $^4$He and a $p$-shell nucleus
it is not obvious which HO frequency is the optimal one due to differences
in the radii of the participating nuclei. In general, the HO frequency in the NCSM is 
typically fixed so that the binding energy has the least dependence on it.
However, while for $^4$He the fastest NCSM convergence and the least dependence 
on the HO frequency is obtained with a higher HO frequency, e.g. $\hbar\Omega>20$ MeV,
for the $p$-shell nuclei the optimal frequency lies typically in the range of 
$\hbar\Omega=10-15$ MeV. In Fig. \ref{he5_he4_omega}, we present the channel cluster
form factor dependence on the HO frequency using a wide range of frequencies:
$\hbar\Omega=13-19$ MeV. It is satisfying that the sensitivity of the 
cluster form factor to the choice of the HO frequency is rather small.

\begin{figure}
\vspace*{2cm}
\includegraphics[width=7.0in]{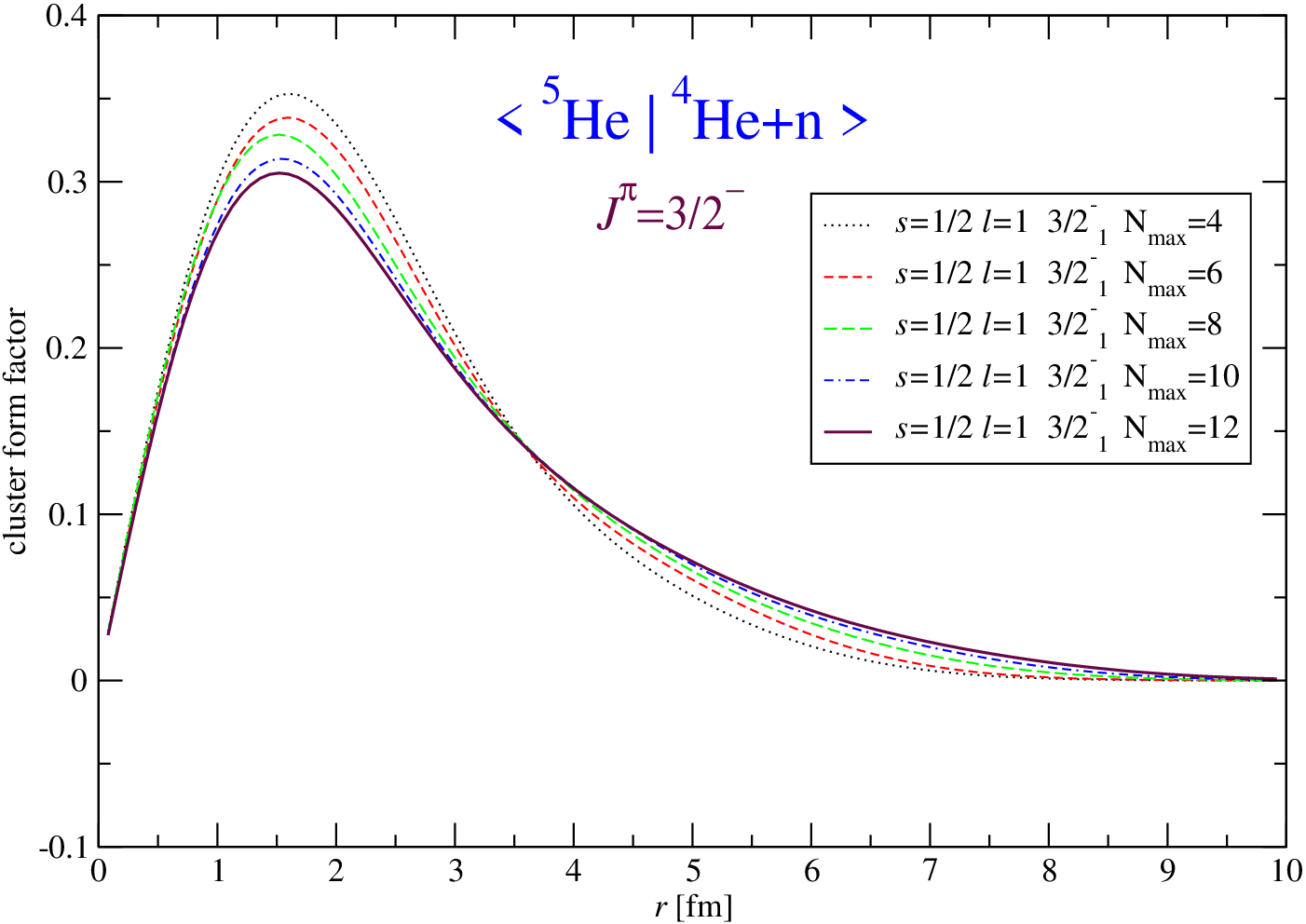}
\caption{\label{he5_he4_nmax} (Color online)
Overlap integral of the $^5$He $\frac{3}{2}^-$ ground state with the
$^4$He+n as a function of separation between the $^4$He and the neutron.
Dependence on the basis size for $N_{\rm max}=4,6,8,10,12$ is presented.
The CD-Bonn 2000 NN potential and the HO frequency of $\hbar\Omega=16$ MeV
were used. The $s$ and $l$ are the channel spin and the relative
angular momentum, respectively. 
}
\end{figure}

\begin{figure}
\vspace*{2cm}
\includegraphics[width=7.0in]{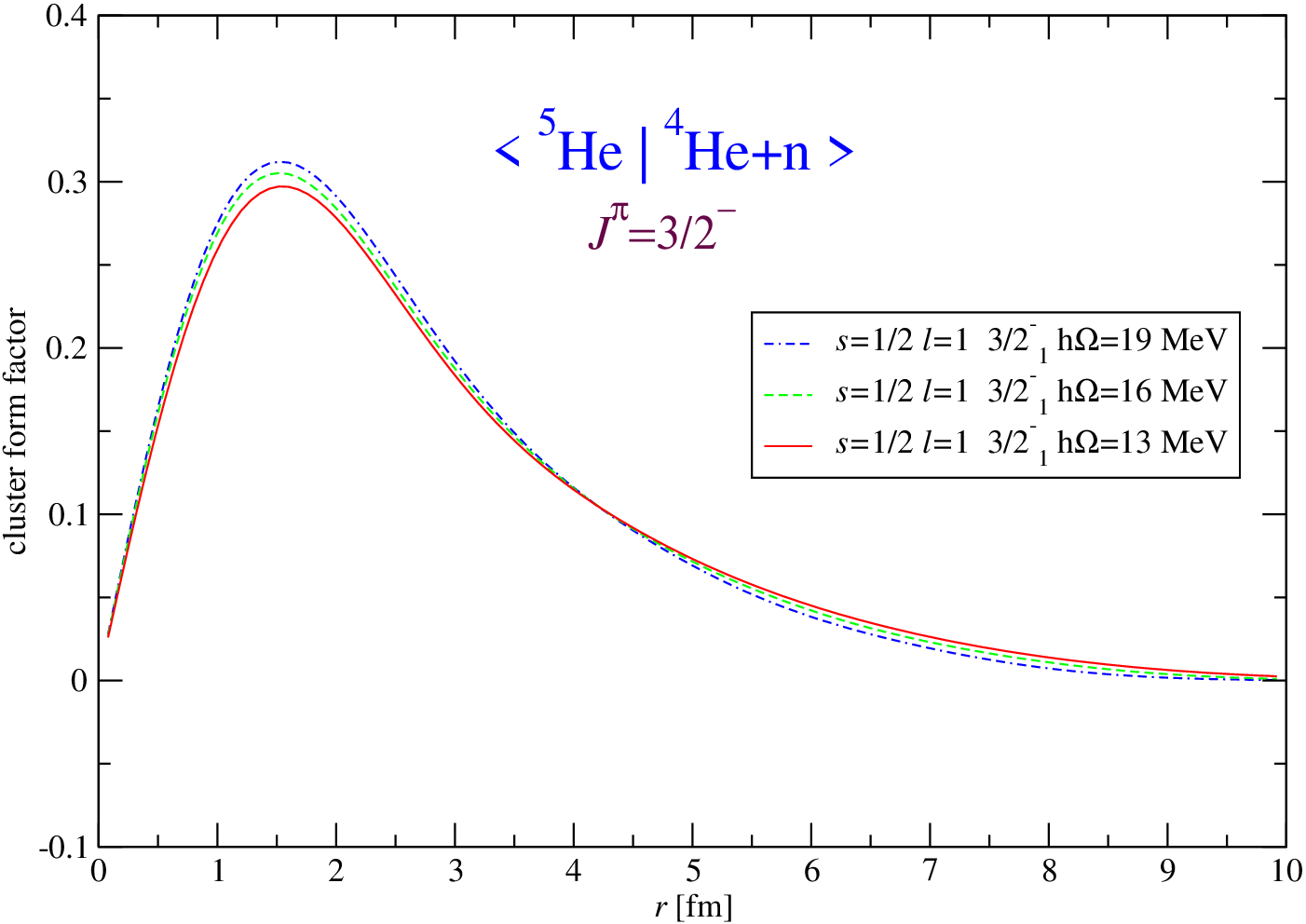}
\caption{\label{he5_he4_omega} (Color online)
Overlap integral of the $^5$He $\frac{3}{2}^-$ ground state with the
$^4$He+n as a function of separation between the $^4$He and the neutron.
Dependence on the HO frequency for $\hbar\Omega=13,16,19$ MeV is presented.
The CD-Bonn 2000 NN potential and the basis size of $N_{\rm max}=12$
were used.
}
\end{figure}

The same as in Figs. \ref{he5_he4_nmax} and \ref{he5_he4_omega} is presented 
in Figs. \ref{he5_he4_nmax_1m} and \ref{he5_he4_omega_1m}
for the excited $\frac{1}{2}^-$ $^5$He resonance. This resonance is broader
in experiment \cite{Till02}. Our calculation shows a more extended overlap integral
for the $\frac{1}{2}^-$ state compared to the $\frac{3}{2}^-$ state.
At the same time, both the basis size and the HO frequency dependencies
are more pronounced. Nevertheless, even for this broad state the conclusions
reached for the $\frac{3}{2}^-$ state apply, which is an encouraging result.

It can be seen in both Figs. \ref{he5_he4_nmax} and \ref{he5_he4_nmax_1m}
how with the increasing $N_{\rm max}$ the overlap extends at large $r$.
However, due to the finiteness of our basis, the overlap integral approaches 
zero with increasing $r$ even for states that correspond to physical resonances.

\begin{figure}
\vspace*{2cm}
\includegraphics[width=7.0in]{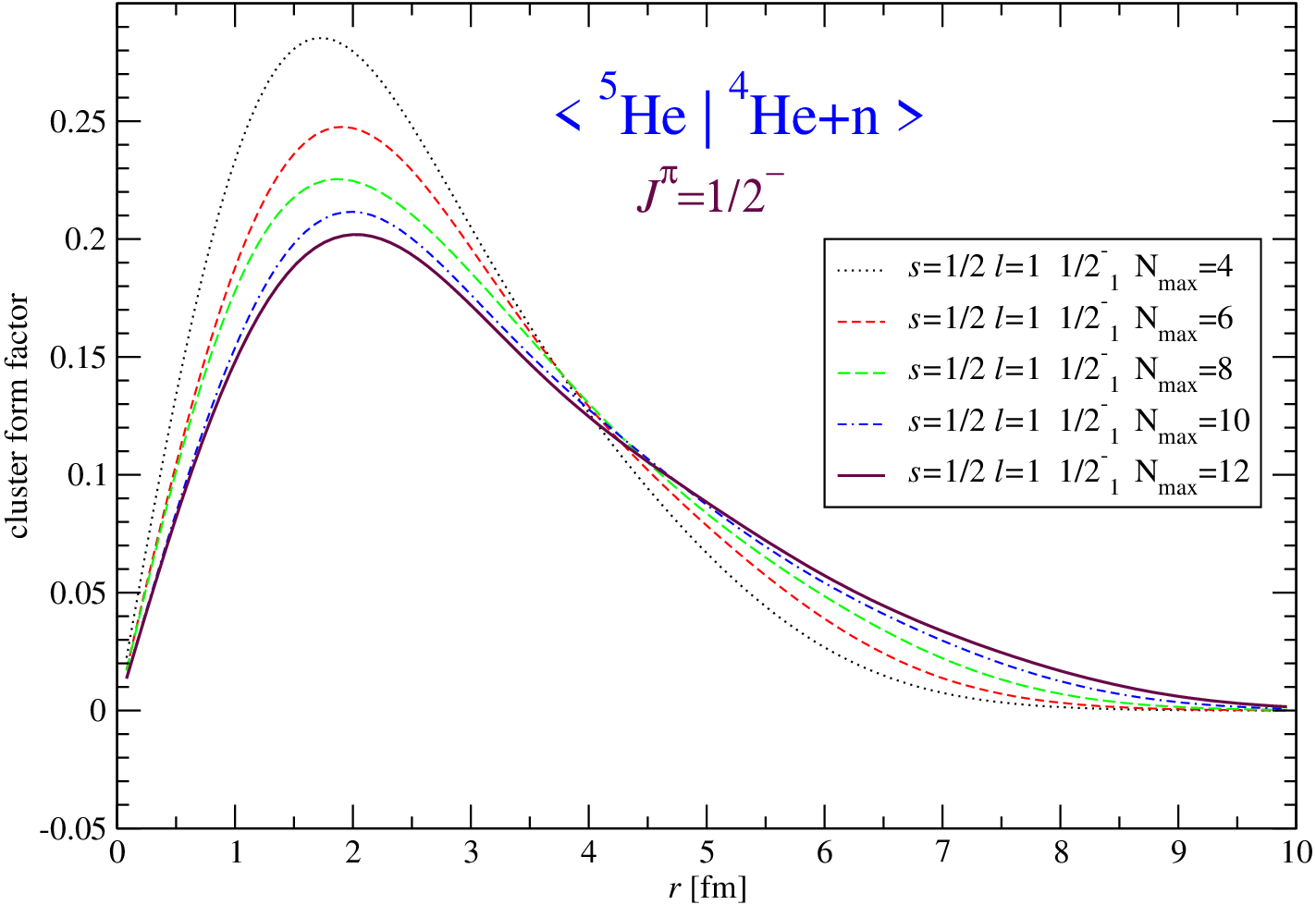}
\caption{\label{he5_he4_nmax_1m} (Color online)
Overlap integral of the $^5$He $\frac{1}{2}^-$ first excited state with the
$^4$He+n as a function of separation between the $^4$He and the neutron.
Dependence on the basis size for $N_{\rm max}=4,6,8,10,12$ is presented.
The CD-Bonn 2000 NN potential and the HO frequency of $\hbar\Omega=16$ MeV
were used.
}
\end{figure}

\begin{figure}
\vspace*{2cm}
\includegraphics[width=7.0in]{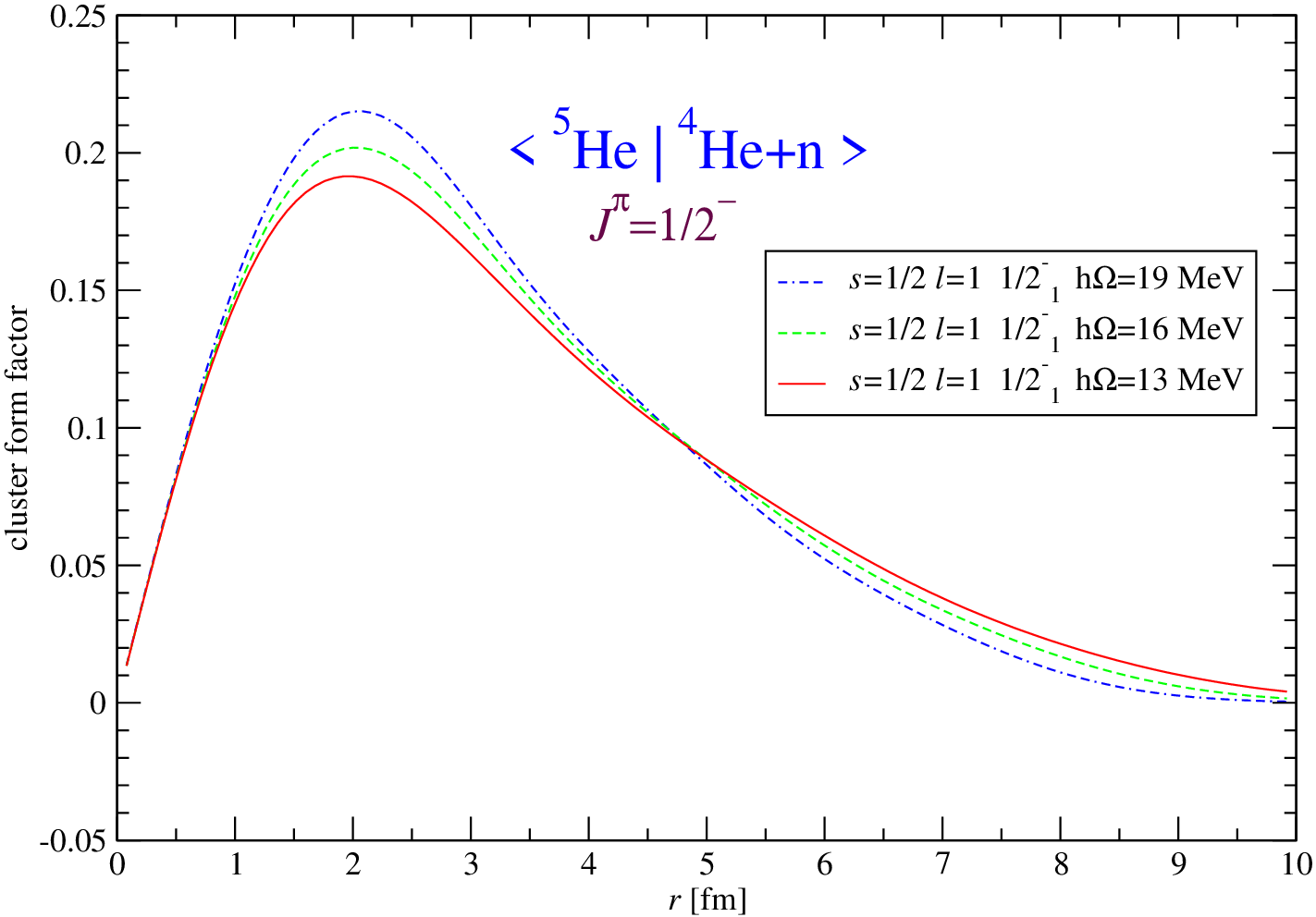}
\caption{\label{he5_he4_omega_1m} (Color online)
Overlap integral of the $^5$He $\frac{1}{2}^-$ first excited state with the
$^4$He+n as a function of separation between the $^4$He and the neutron.
Dependence on the HO frequency for $\hbar\Omega=13,16,19$ MeV is presented.
The CD-Bonn 2000 NN potential and the basis size of $N_{\rm max}=12$
were used.
}
\end{figure}

\subsection{$\langle ^5$He$|$t+d$\rangle$}\label{he5_h3}

In Fig. \ref{he5_h3_omega}, we present the NCSM calculated channel cluster
form factors for the $\langle ^5$He$|$t+d$\rangle$ $J^\pi=\frac{3}{2}^+$ resonance.
The NCSM calculations were performed using the CD-Bonn 2000 NN potential 
in the $11\hbar\Omega$ model space for $^5$He.
There are three possible channels corresponding to the combinations
of the channel spin $s=\frac{3}{2}$ and $s=\frac{1}{2}$ and the relative orbital
momenta $l=0$ and $l=2$. Clearly, the overlap integral is by far the largest
for the $s$-wave channel $s=\frac{3}{2}$ $l=0$. The $d$-wave channels give
small overlap integrals with the $s=\frac{3}{2}$ $l=2$ channel greater
than the $s=\frac{1}{2}$ $l=2$ channel. Results corresponding to three different
HO frequencies are shown in Fig. \ref{he5_h3_omega}. Despite the wide range
of the HO frequencies, changes in the resulting overlap integrals are not 
significant. 

\begin{figure}
\vspace*{2cm}
\includegraphics[width=7.0in]{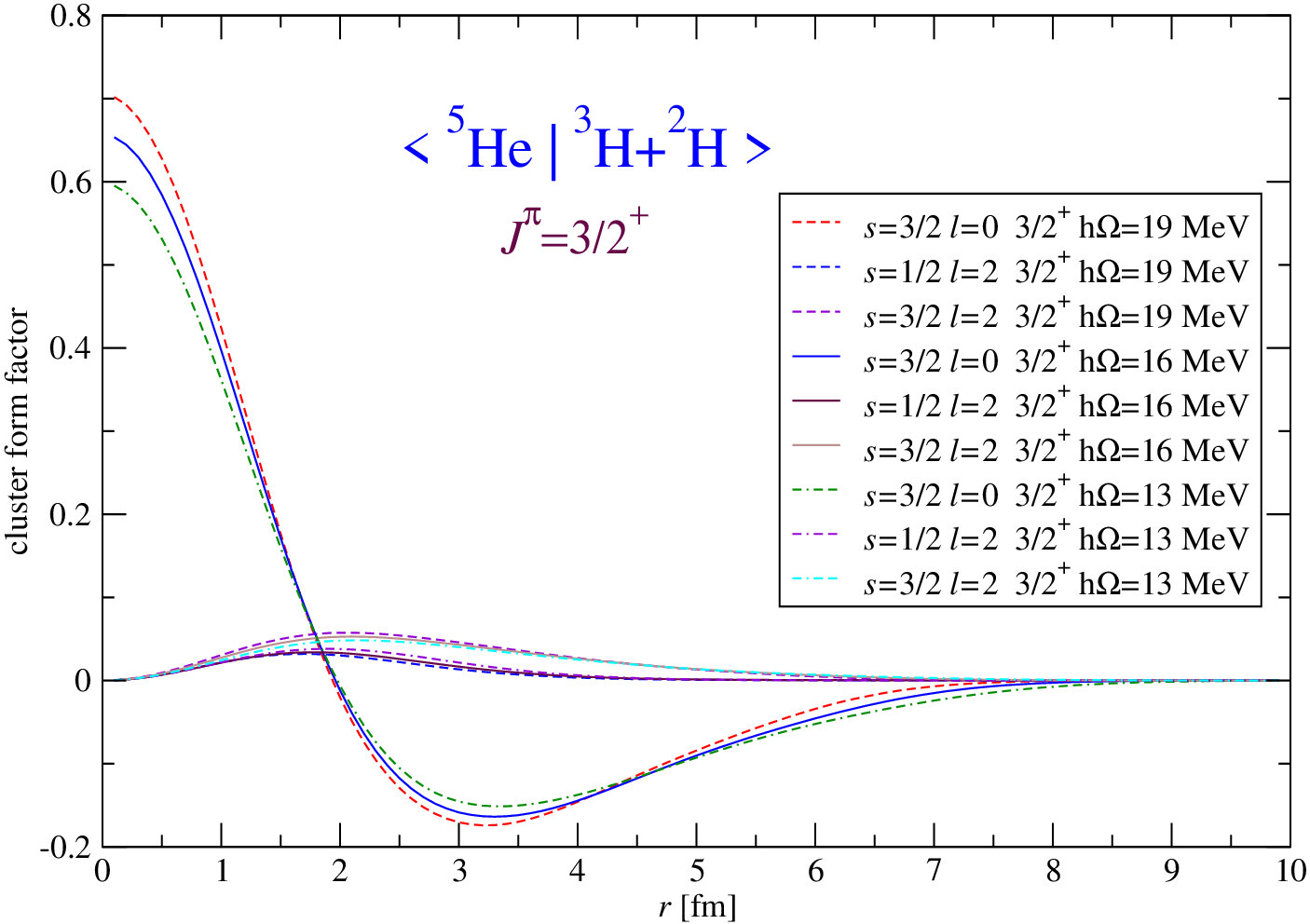}
\caption{\label{he5_h3_omega} (Color online)
Overlap integral of the $^5$He $\frac{3}{2}^+$ excited state with the
$^3$H+d as a function of separation between the $^3$H and the deuteron.
Dependence on the HO frequency for $\hbar\Omega=13,16,19$ MeV is presented.
The CD-Bonn 2000 NN potential and the basis size of $N_{\rm max}=12$ (for $^3$H and d)
$N_{\rm max}=11$ (for $^5$He) were used.
}
\end{figure}

It is interesting to point out that the d+t $\frac{3}{2}^+$ resonance in $^5$He
appears as the second $\frac{3}{2}^+$ state in the NCSM calculations 
reported in this paper as well as in the previously published $^5$He results 
\cite{ZVB94,start_en_indp_NCSM}. The appearance of low-lying positive parity states
in the NCSM calculations of the $^5$He spectrum was criticized in Ref. \cite{CL96}.
We note that such states were also observed in phenomenological shell model calculations
\cite{HG83,Popp93}. At the same time, some evidence for a low-lying $\frac{1}{2}^+$ state
was reported in the R-matrix analysis of Ref. \cite{W88}. No such state was, however,
included in the recent evaluation \cite{Till02}. The low-lying $\frac{3}{2}^+$ state
obtained in the present calculation has basically zero overlap with the d+t. It is quite
possible that it corresponds to a non-resonant continuum state of a free neutron and
the $^4$He that appears as an excited state due to the finiteness of the basis used 
in our investigation. The d+t resonant $\frac{3}{2}^+$ state is dominated by the $s^3p^2$ 
configuration. 


\subsection{$\langle ^6$Li$| ^4$He+d$\rangle$}\label{li6_he4}

Our $\langle ^6$Li$| ^4$He+d$\rangle$ channel cluster form factors for the $^6$Li
$J^\pi=1^+,3^+$ and $2^+$ are presented in Fig. \ref{li6_he4_nmax_1p}, 
Fig. \ref{li6_he4_nmax_3p} and Fig. \ref{li6_he4_nmax_2p}, respectively. 
The corresponding
spectroscopic factors are then summarized in Table \ref{tab_li6_he4}.
In the NCSM calculations, we used the $^6$Li wave functions obtained in Ref. 
\cite{NCSM_6} using the CD-Bonn NN potential \cite{Machl}. In the three figures, 
the thick lines correspond to the $10\hbar\Omega$ results and the thin
lines to the $8\hbar\Omega$ $^6$Li results. We can see only small changes in the
overlap integrals when the basis size is changed, in particular for the
$1^+ 0$ ground state and the $3^+ 0$ excited state. It is interesting to note 
that the ground state is dominated by the $s=1$, $l=0$ $^4$He+d configuration
while the excited $3^+ 0$, $2^+0$ and $1^+_2 0$ states are dominated
by the $s=1$, $l=2$ $^4$He+d configuration. This is in agreement with the
analysis of the $^4$He+d elastic scattering experimental data \cite{Till02}.

\begin{figure}
\vspace*{2cm}
\includegraphics[width=7.0in]{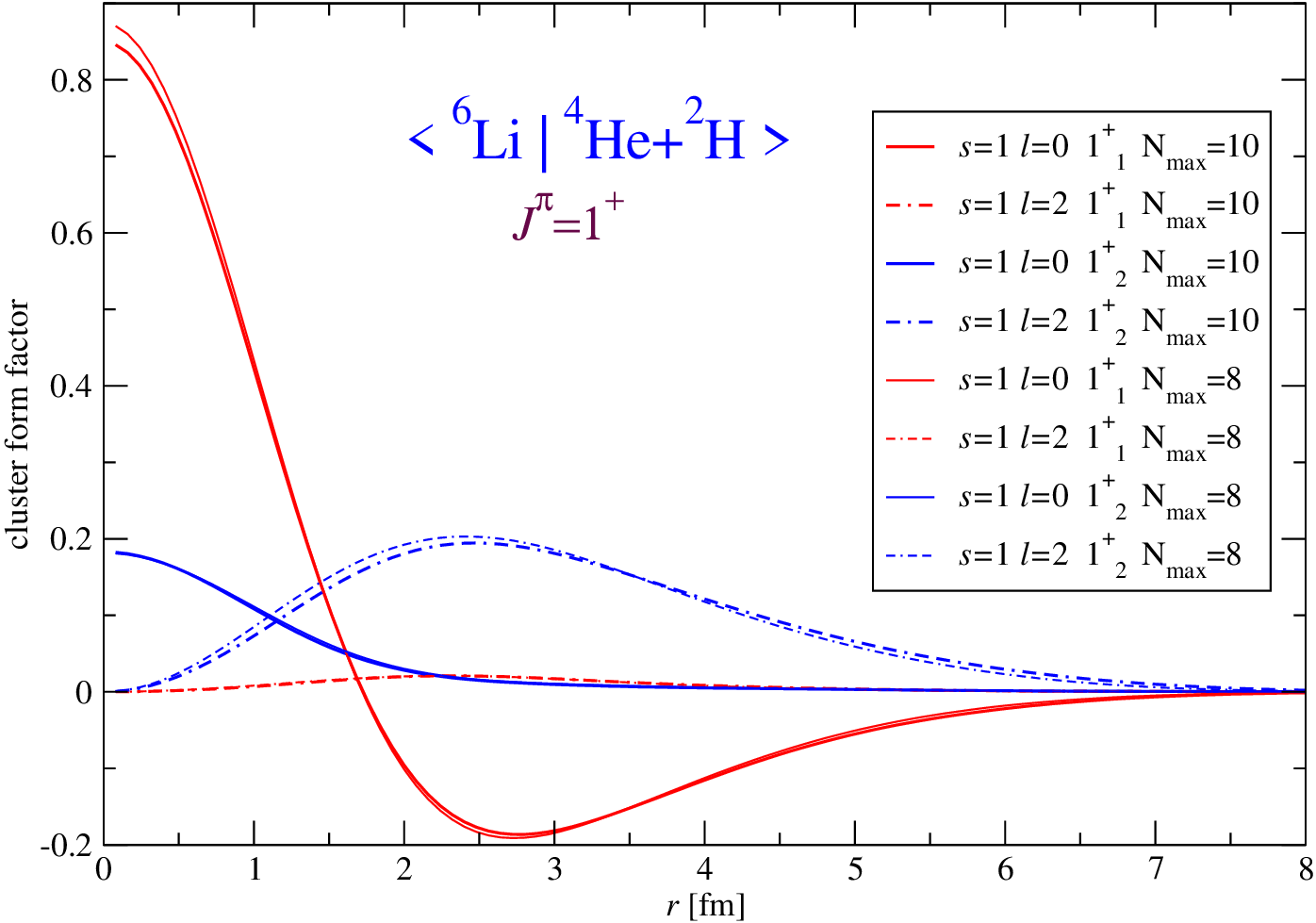}
\caption{\label{li6_he4_nmax_1p} (Color online)
Overlap integral of the $^6$Li $1^+_1 0$ ground state and the $1^+_2 0$ first excited 
with the $^4$He+d as a function of separation between the $^4$He and the deuteron.
Dependence on the basis size for $N_{\rm max}=8,10$ is presented.
The CD-Bonn NN potential and the HO frequency of $\hbar\Omega=13$ MeV
were used.
}
\end{figure}

\begin{figure}
\vspace*{2cm}
\includegraphics[width=7.0in]{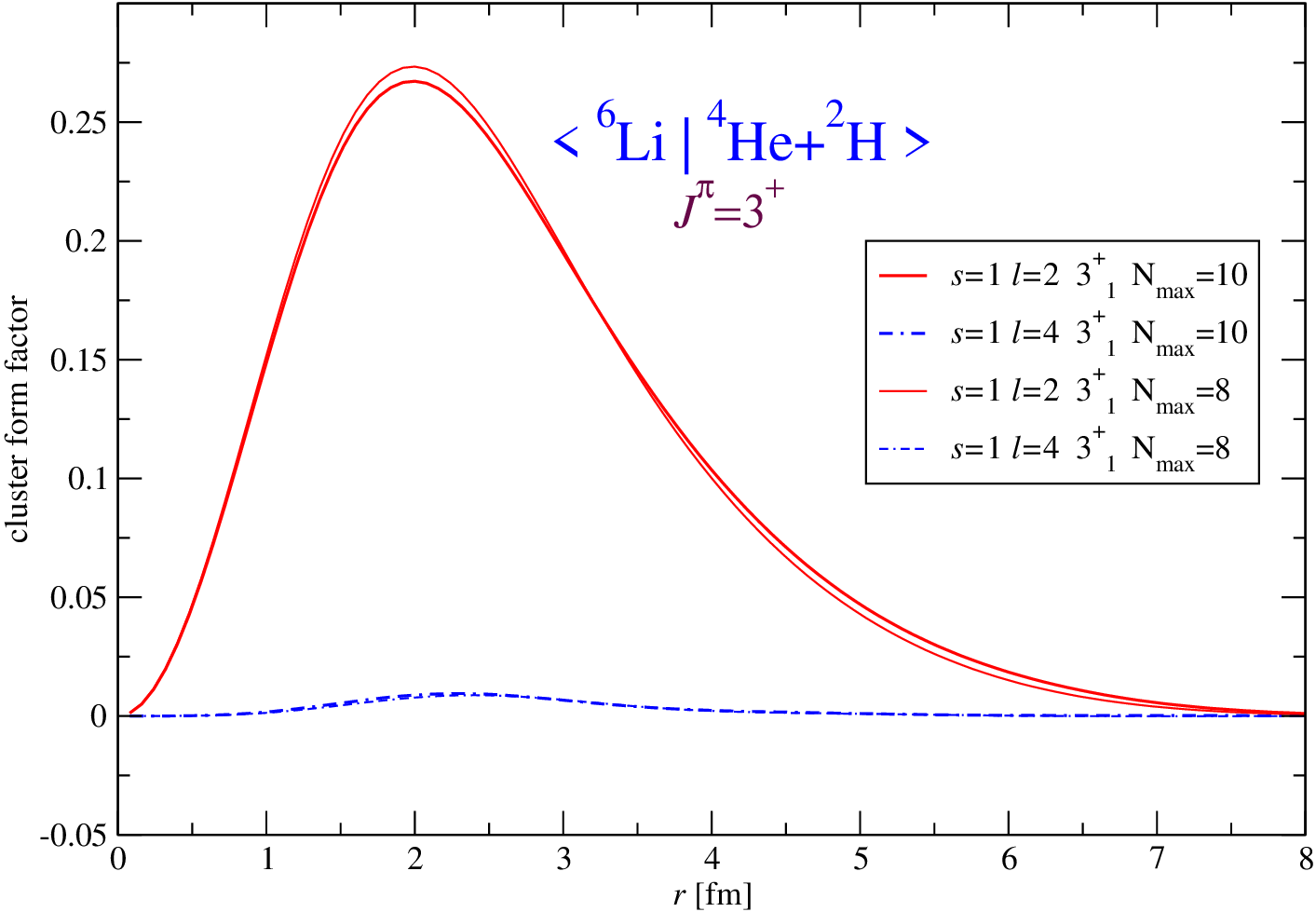}
\caption{\label{li6_he4_nmax_3p} (Color online)
Overlap integral of the $^6$Li $3^+ 0$ first excited state with the
$^4$He+d as a function of separation between the $^4$He and the deuteron.
Dependence on the basis size for $N_{\rm max}=8,10$ is presented.
The CD-Bonn NN potential and the HO frequency of $\hbar\Omega=13$ MeV
were used.
}
\end{figure}

\begin{figure}
\vspace*{2cm}
\includegraphics[width=7.0in]{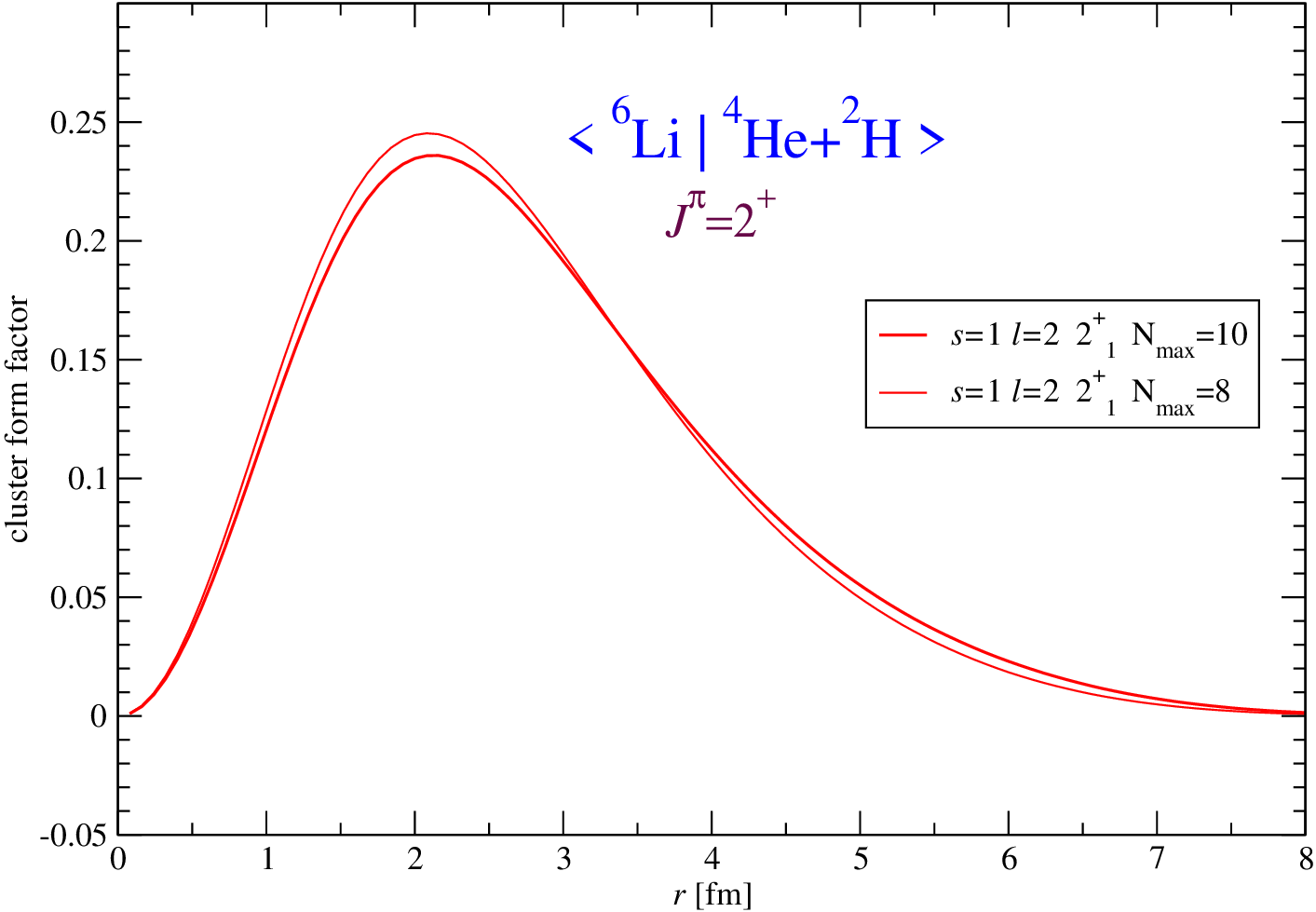}
\caption{\label{li6_he4_nmax_2p} (Color online)
Overlap integral of the $^6$Li $2^+ 0$ excited state with the
$^4$He+d as a function of separation between the $^4$He and the deuteron.
Dependence on the basis size for $N_{\rm max}=8,10$ is presented.
The CD-Bonn NN potential and the HO frequency of $\hbar\Omega=13$ MeV were used.
}
\end{figure}

\begin{table}
\begin{tabular}{c| c c| c c}
\multicolumn{5} {c} {$\langle ^6$Li$| ^4$He+d$\rangle$} \\
\hline
$J^\pi T$ & $(s,l)$ & $S$  & $(s,l)$ & $S$   \\
\hline
$1^+_1 0$ & $(1,0)$ & 0.822 & $(1,2)$ & 0.006  \\
$3^+_1 0$ & $(1,2)$ & 0.890 & $(1,4)$ & 0.0008 \\
$2^+_1 0$ & $(1,2)$ & 0.864 &         &  \\
$1^+_2 0$ & $(1,0)$ & 0.017 & $(1,2)$ & 0.811 \\
$1^+_3 0$ & $(1,0)$ & 0.031 & $(1,2)$ & 0.088 \\
\end{tabular}
\caption{\label{tab_li6_he4} Spectroscopic factors for the $\langle ^6$Li$| ^4$He+d$\rangle$
corresponding to the $^6$Li ground and excited states and the $^4$He ground state. 
The CD-Bonn NN potential, the basis size of $N_{\rm max}=10$ for $^6$Li
and $N_{\rm max}=12$ (for $^4$He and d) and the HO frequency of $\hbar\Omega=13$ MeV
were used. The $s$ and $l$ are the channel spin and the relative
angular momentum, respectively. 
}
\end{table}

\subsection{$\langle ^6$Be$| ^3$He+$^3$He$\rangle$}\label{be6_he3}

Just as a resonance plays a critical role in the
rate of the d+t reaction, there is some speculation that the
reaction $^3$He($^3$He,2p)$^4$He, which is important to the Standard Solar Model (SSM),
could be affected by a resonance in the composite $^6$Be
system~\cite{r:6Be_res}. Although recent experiments at the LUNA underground 
facility~\cite{r:LUNA} seem not to favor a narrow resonance, 
they do not definitively rule out its presence~\cite{r:6Be_res}. To investigate 
a possibility of a resonance in the $^3$He+$^3$He system, we performed NCSM
calculations of the overlap integrals $\langle ^6$Be$| ^3$He+$^3$He$\rangle$ for the
lowest four $0^+ 1$ states obtained in the NCSM description of $^6$Be.
The calculations were performed using the CD-Bonn 2000 NN potential in the basis
spaces up to $10\hbar\Omega$ for $^6$Be. The lowest two $0^+ 1$ states are the
$p$-shell dominated states while the third and the fourth $0^+ 1$ state is a
one-particle-one-hole and a two-particle-two-hole dominated state, respectively.
In the $N_{\rm max}=10$ ($10\hbar\Omega$) basis space and the $\hbar\Omega=13$ MeV
calculation, their excitation energy is 12.5 MeV and 13.5 MeV, respectively, 
not far from the $^3$He+$^3$He threshold. However, the excitation energy of these states
is not yet converged in the present calculation and it is expected to further
decrease with the basis size enlargement. Our channel cluster form factor results
obtained in the $10\hbar\Omega$ space are shown in Fig. \ref{be6_he3_nmax}. The dependence
on the HO frequency is presented for all four states. A large overlap integral
is found for the ground state and also for the one-particle-one-hole dominated $0^+_3 1$ 
state. On the other hand, the overlap integral for the $0^+_2 1$ state is negligible
and the one for the $0^+_4 1$ is quite small. It is interesting to note a stronger
HO frequency dependence of the overlap integrals for the $2\hbar\Omega$ dominated states
compared to the $p$-shell states. This is another manifestation of a slower convergence
of these states in the NCSM. The significant overlap integral of the $0^+_3 1$ $^6$Be
state suggest that this state might contribute as a resonance
in the $^3$He+$^3$He reaction. However, our prediction of its excitation energy 
is not certain. Based on our NCSM results up to $N_{\rm max}=10$ we expect this state
to converge below the $^3$He+$^3$He threshold of 11.49 MeV.  

\begin{figure}
\vspace*{2cm}
\includegraphics[width=7.0in]{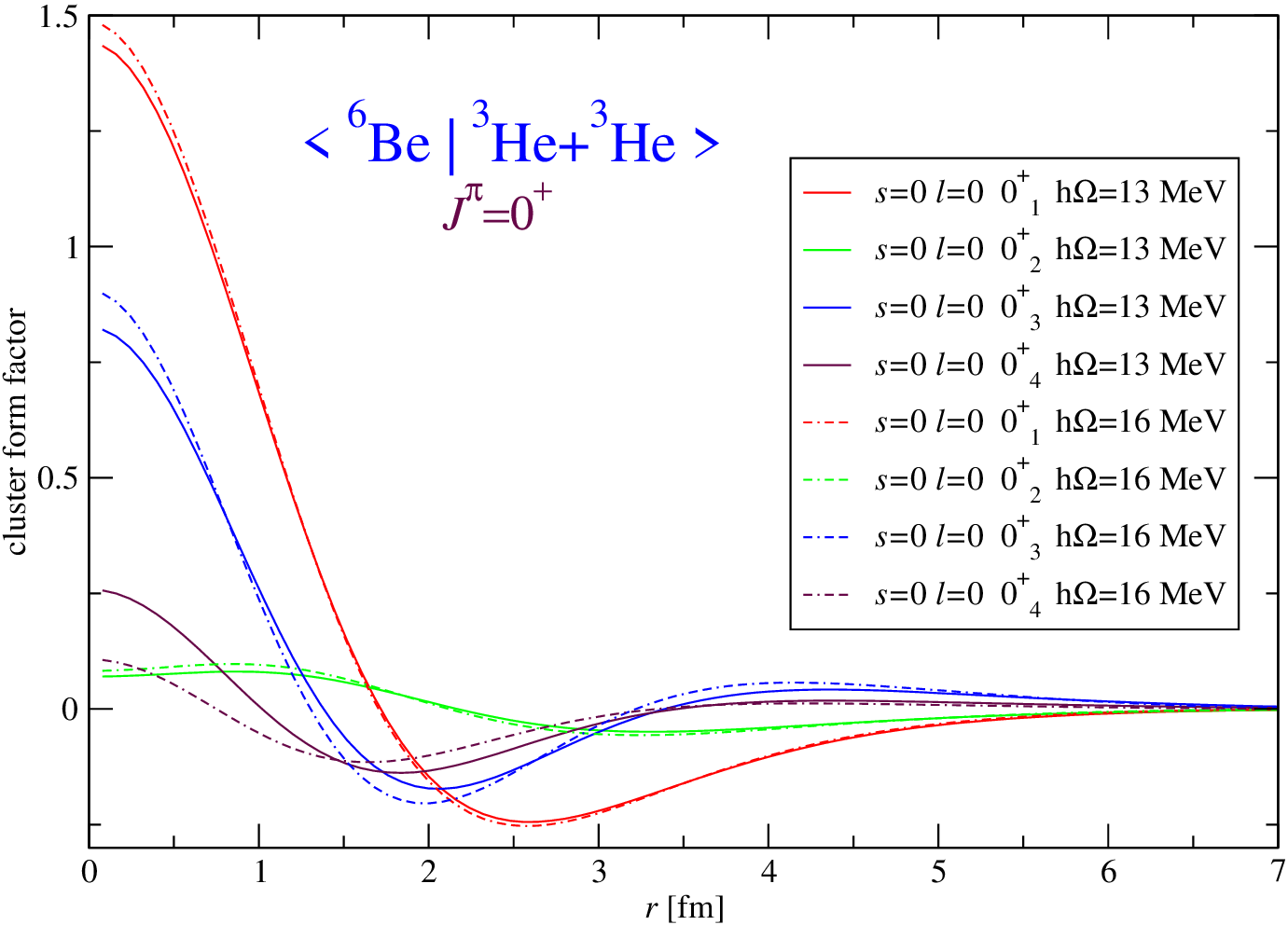}
\caption{\label{be6_he3_nmax} (Color online)
Overlap integral of the $^6$Be $0^+ 0$ states with the
$^3$He+$^3$He as a function of separation between the $^3$He nuclei.
Dependence on the HO frequency for $\hbar\Omega=13,16$ MeV is presented.
The CD-Bonn 2000 NN potential and the basis size of $N_{\rm max}=10$ (for $^6$Be),
$N_{\rm max}=12$ (for $^3$He) were used. The $0^+_1 1$ and $0^+_2 1$ are 
$p$-shell states. The $0^+_3 1$ and $0^+_4 1$ are a one-particle-one-hole dominated
and a two-particle-two-hole dominated state, respectively. 
}
\end{figure}

\subsection{$\langle ^7$Li$| ^4$He+t$\rangle$}\label{li7_he4}

Our results for the $\langle ^7$Li$| ^4$He+t$\rangle$ channel cluster form factors 
are shown in Fig. \ref{li7_he4_J}, while the corresponding spectroscopic factors 
are summarized in Table \ref{tab_li7_he4}. Apart from the large overlap integrals and 
spectroscopic factors for the bound $\frac{3}{2}^-_1$ and $\frac{1}{2}^-_1$ states
we find these quantities large also for the first excited $\frac{7}{2}^-_1$
and the first excited $\frac{5}{2}^-_1$ state. Both these states appear as resonances
in the $^4$He+t cross section \cite{Till02}. The present results can be compared
to the three-nucleon transfer calculations of Ref. \cite{KM_3N_spec} 
obtained using the phenomeological Cohen-Kurath interaction \cite{CK}. 
The agreement for the lowest four states is quite good. 
For the second excited $\frac{5}{2}^-_2$ state, however,
our spectroscopic factor is significantly smaller than the one obtained 
in Ref. \cite{KM_3N_spec}.

\begin{figure}
\vspace*{2cm}
\includegraphics[width=7.0in]{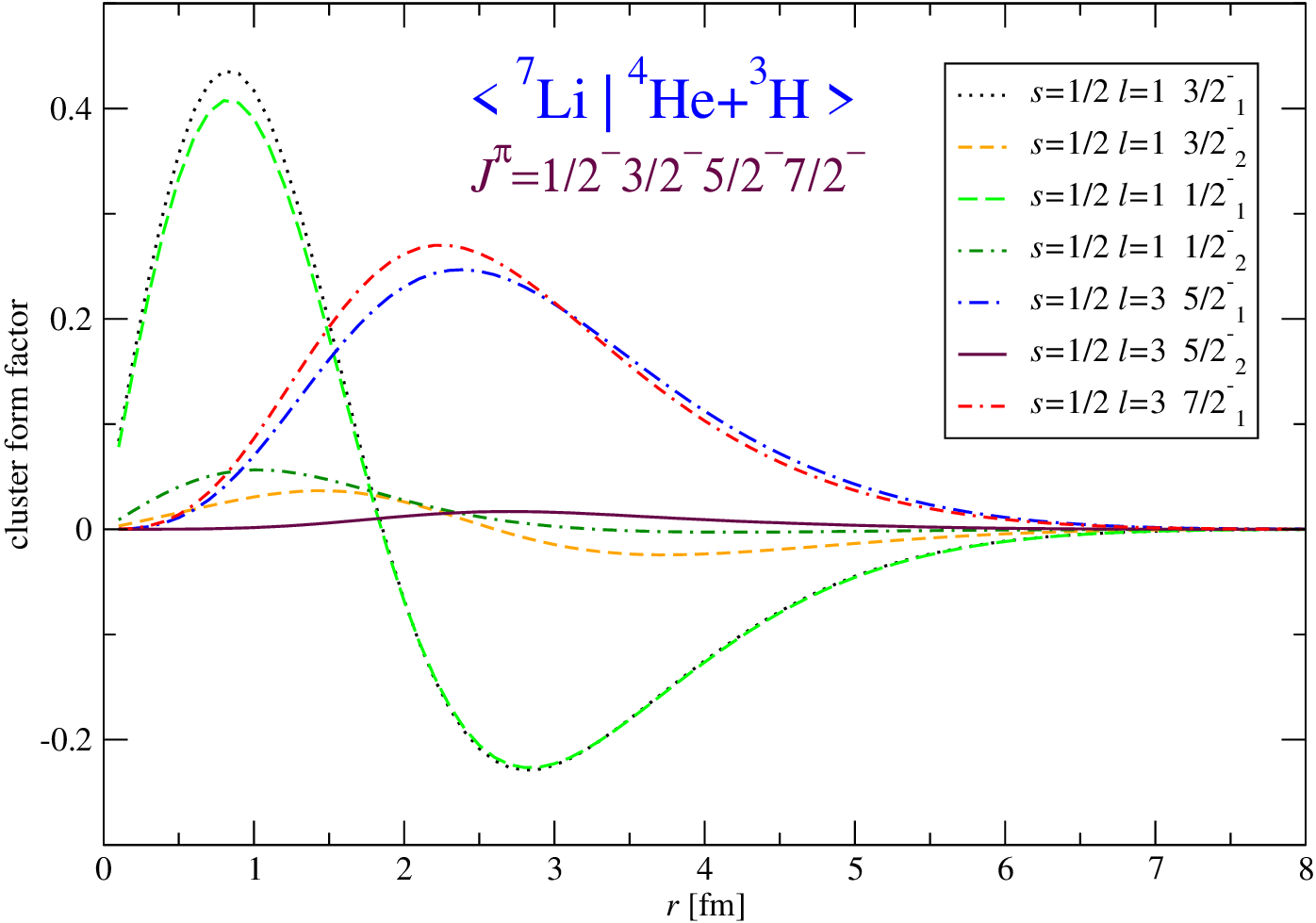}
\caption{\label{li7_he4_J} (Color online)
Overlap integral of the $^7$Li low-lying 
$J=\frac{1}{2}^-,\frac{3}{2}^-,\frac{5}{2}^-,\frac{7}{2}^-$ states with the
$^4$He+$^3$H as a function of separation between the $^4$He and the triton.
The CD-Bonn 2000 NN potential, the basis size of $N_{\rm max}=8$ (for $^7$Li),
$N_{\rm max}=10$ (for $^4$He and $^3$H)  and the HO frequency 
of $\hbar\Omega=13$ MeV were used.
}
\end{figure}

\begin{table}
\begin{tabular}{c| c c}
\multicolumn{3} {c} {$\langle ^7$Li$| ^4$He+$^3$H$\rangle$} \\
\hline
$J^\pi T$ & $(s,l)$ & $S$  \\
\hline
$\frac{3}{2}^-_1 \frac{1}{2}$ & $(\frac{1}{2},1)$ & 0.941 \\
$\frac{1}{2}^-_1 \frac{1}{2}$ & $(\frac{1}{2},1)$ & 0.923 \\
$\frac{7}{2}^-_1 \frac{1}{2}$ & $(\frac{1}{2},3)$ & 0.906 \\
$\frac{5}{2}^-_1 \frac{1}{2}$ & $(\frac{1}{2},3)$ & 0.883 \\
$\frac{5}{2}^-_2 \frac{1}{2}$ & $(\frac{1}{2},3)$ & 0.005 \\
$\frac{3}{2}^-_2 \frac{1}{2}$ & $(\frac{1}{2},1)$ & 0.020  \\
$\frac{1}{2}^-_2 \frac{1}{2}$ & $(\frac{1}{2},1)$ & 0.007 \\
$\frac{7}{2}^-_2 \frac{1}{2}$ & $(\frac{1}{2},3)$ & 0.056 \\
$\frac{5}{2}^-_3 \frac{1}{2}$ & $(\frac{1}{2},3)$ & 0.013 \\
$\frac{1}{2}^-_3 \frac{1}{2}$ & $(\frac{1}{2},1)$ & 0.036 \\
$\frac{5}{2}^-_4 \frac{1}{2}$ & $(\frac{1}{2},3)$ & 0.064 \\
\end{tabular}
\caption{\label{tab_li7_he4} Spectroscopic factors for the $\langle ^7$Li$| ^4$He+$^3$H$\rangle$
corresponding to the $^7$Li ground and excited states and the $^4$He ground state. 
The CD-Bonn 2000 NN potential, the basis size of $N_{\rm max}=8$ (for $^7$Li),
$N_{\rm max}=10$ (for $^4$He and $^3$H) and the HO frequency 
of $\hbar\Omega=13$ MeV were used. The $s$ and $l$ are the channel spin and the relative
angular momentum, respectively. 
}
\end{table}

\subsection{$\langle ^7$Li$| ^6$Li+n$\rangle$}\label{li7_li6}

The other system involving $^7$Li as the composite nucleus that we investigated
is $^6$Li+n. Our calculated overlap integrals are summarized in Fig. \ref{li7_li6_J}.
The corresponding spectroscopic factors are given in Table \ref{tab_li7_li6}.
As in the $\langle ^7$Li$| ^4$He+t$\rangle$ case, we observe large overlap integrals
and spectroscopic factors for the two bound states $\frac{3}{2}^-_1$ and $\frac{1}{2}^-_1$.
Contrary to the $\langle ^7$Li$| ^4$He+t$\rangle$ case, however, we find a large overlap integral
and the spectroscopic factor for the $\frac{5}{2}^-_2$ state. The lowest
$\frac{7}{2}^-_1$ and $\frac{5}{2}^-_1$ states have negligible overlap integrals
for the $^6$Li+n system. The large overlap integral
and the spectroscopic factor for the $\frac{5}{2}^-_2$ state is consistent with
the observed resonance in the $^6$Li+n cross section. In addition to the $\frac{5}{2}^-_2$
state, we also find large overlap integrals for the higher lying $\frac{3}{2}^-_2$
and $\frac{1}{2}^-_2$ states. In Fig. \ref{li7_li6_5m}, we display the basis size dependence
of the $\frac{5}{2}^-_1$ and the $\frac{5}{2}^-_2$ states for the $N_{\rm max}=4,6$ and 8 
calculations. The results for the resonant $\frac{5}{2}^-_2$ state are fairly robust.
The spectroscopic factor of the $\frac{5}{2}^-_1$ state show a stronger basis size 
dependence. Our $N_{\rm max}=8$ spectroscopic factor for the
$\frac{5}{2}^-_1$ state is 0.016 for $s=\frac{3}{2}$, $l=1$ channel, 
a significantly smaller value than that obtained using the Cohen-Kurath $0p$-shell 
phenomenological interaction \cite{CK_spec}. Our NCSM result is, however, consistent with the
spectroscopic factor obtained using the variational Monte Carlo wave functions \cite{Wiringa_pr}. 

\begin{figure}
\vspace*{2cm}
\includegraphics[width=7.0in]{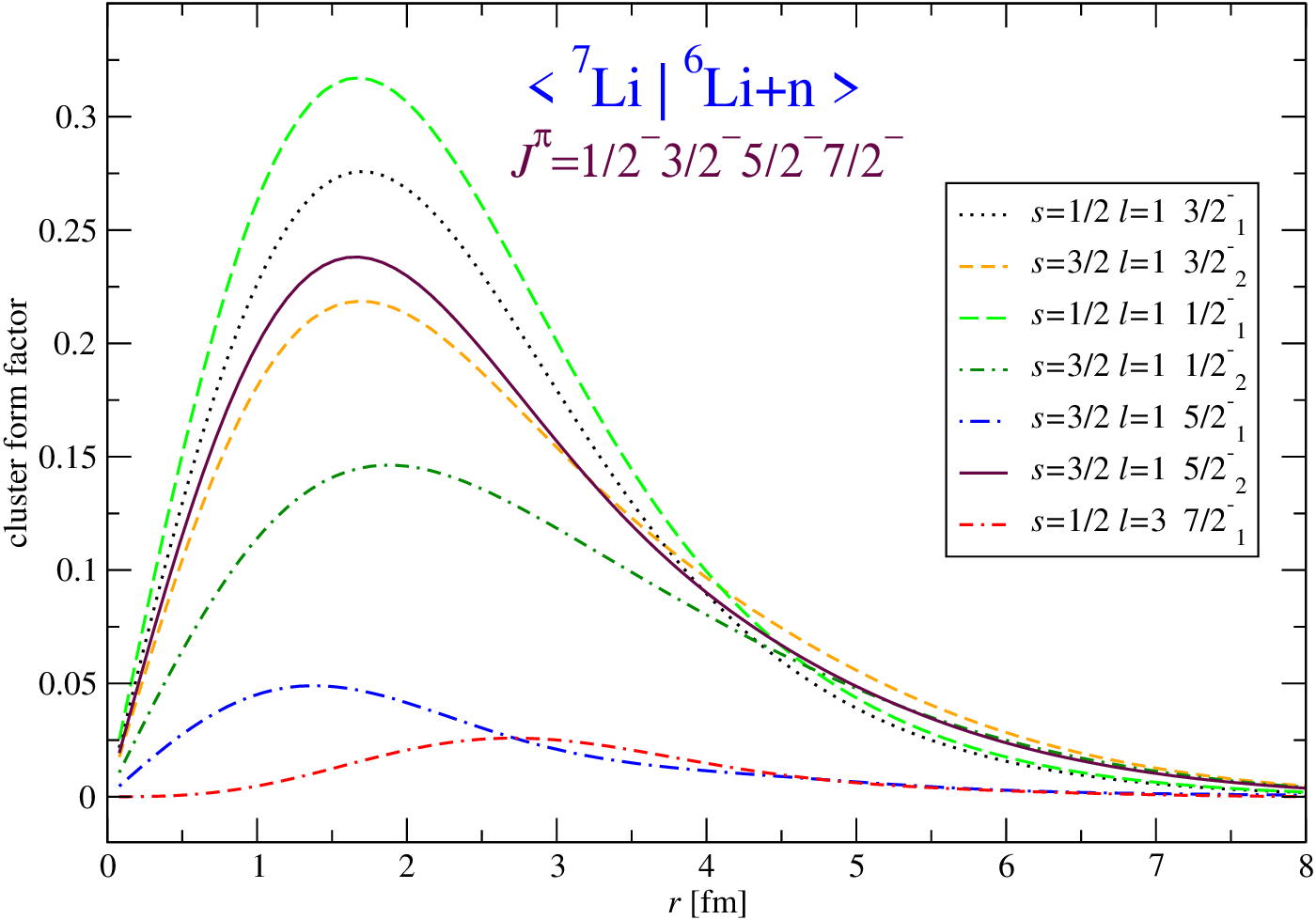}
\caption{\label{li7_li6_J} (Color online)
Overlap integral of the $^7$Li low-lying 
$J=\frac{1}{2}^-,\frac{3}{2}^-,\frac{5}{2}^-,\frac{7}{2}^-$ states with the
$^6$Li+n as a function of separation between the $^6$Li and the neutron.
The CD-Bonn 2000 NN potential, the basis size of $N_{\rm max}=8$ and the HO frequency 
of $\hbar\Omega=13$ MeV were used.
}
\end{figure}

\begin{figure}
\vspace*{2cm}
\includegraphics[width=7.0in]{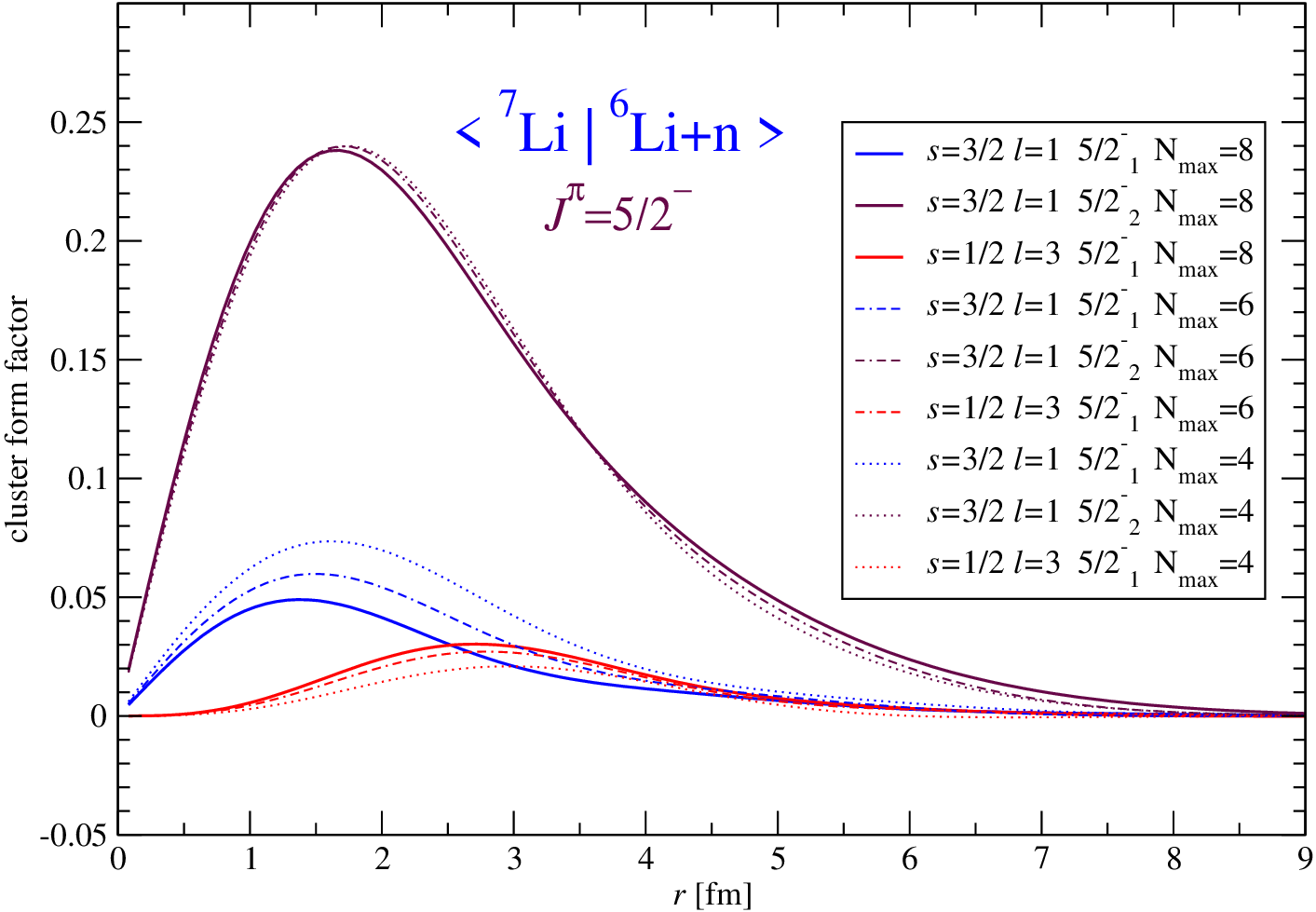}
\caption{\label{li7_li6_5m} (Color online)
Overlap integral of the $^7$Li low-lying $\frac{5}{2}^-$ states with the
$^6$Li+n as a function of separation between the $^6$Li and the neutron.
Dependence on the basis size for $N_{\rm max}=4,6,8$ is presented.
The CD-Bonn 2000 NN potential and the HO frequency 
of $\hbar\Omega=13$ MeV were used.
}
\end{figure}

\begin{table}
\begin{tabular}{c| c c| c c| c c}
\multicolumn{7} {c} {$\langle ^7$Li$| ^6$Li+n$\rangle$} \\
\hline
$J^\pi T$ & $(s,l)$ & $S$  & $(s,l)$ & $S$ & $(s,l)$ & $S$   \\
\hline
$\frac{3}{2}^-_1 \frac{1}{2}$ & $(\frac{1}{2},1)$ & 0.806 & $(\frac{3}{2},1)$ & 0.015  
& $(\frac{3}{2},3)$ & 0.002 \\  
$\frac{1}{2}^-_1 \frac{1}{2}$ & $(\frac{1}{2},1)$ & 1.027 & $(\frac{3}{2},1)$ & 0.004 & & \\
$\frac{7}{2}^-_1 \frac{1}{2}$ & $(\frac{1}{2},3)$ & 0.012 & $(\frac{3}{2},3)$ & 0.0001 
& $(\frac{3}{2},5)$ & 0.0005 \\    
$\frac{5}{2}^-_1 \frac{1}{2}$ & $(\frac{3}{2},1)$ & 0.016 & $(\frac{1}{2},3)$ & 0.017  
& $(\frac{3}{2},3)$ & 0.0003 \\    
$\frac{5}{2}^-_2 \frac{1}{2}$ & $(\frac{3}{2},1)$ & 0.688 & $(\frac{1}{2},3)$ & 0.0001 
& $(\frac{3}{2},3)$ & 0.0003 \\    
$\frac{3}{2}^-_2 \frac{1}{2}$ & $(\frac{1}{2},1)$ & 0.005 & $(\frac{3}{2},1)$ & 0.693 
& $(\frac{3}{2},3)$ & 0.0001 \\  
$\frac{1}{2}^-_2 \frac{1}{2}$ & $(\frac{1}{2},1)$ & 0.186  & $(\frac{3}{2},1)$ & 0.414 & & \\
$\frac{7}{2}^-_2 \frac{1}{2}$ & $(\frac{1}{2},3)$ & 0.001 & $(\frac{3}{2},3)$ & 0.0001 
& $(\frac{3}{2},5)$ & 0.0000 \\
$\frac{5}{2}^-_3 \frac{1}{2}$ & $(\frac{3}{2},1)$ & 0.020 & $(\frac{1}{2},3)$ & 0.0003 
& $(\frac{3}{2},3)$ & 0.0007 \\    
$\frac{1}{2}^-_3 \frac{1}{2}$ & $(\frac{1}{2},1)$ & 0.089 & $(\frac{3}{2},1)$ & 0.223 & & \\
$\frac{5}{2}^-_4 \frac{1}{2}$ & $(\frac{3}{2},1)$ & 0.006 & $(\frac{1}{2},3)$ & 0.003 
& $(\frac{3}{2},3)$ & 0.0009 \\    
\end{tabular}
\caption{\label{tab_li7_li6} Spectroscopic factors for the $\langle ^7$Li$| ^6$Li+n$\rangle$
corresponding to the $^7$Li ground and excited states and the $^6$Li ground state. 
The CD-Bonn 2000 NN potential, the basis size of $N_{\rm max}=8$ and the HO frequency 
of $\hbar\Omega=13$ MeV were used. The $s$ and $l$ are the channel spin and the relative
angular momentum, respectively. 
}
\end{table}

\subsection{$\langle ^{8}$Be$| ^{6}$Li+d$\rangle$}\label{be8_li6}

We also investigate systems with $^8$Be as the composite nucleus. 
The $^6$Li+d reactions in particular are of some interest in controlled thermonuclear
research and their cross sections have been measured \cite{McClenahan75}. At the same time, they are 
a part of a reaction network with $^8$Be as the composite nucleus, which is being analyzed 
by the R-matrix method \cite{Page}.
Our calculated spectroscopic factors for the $^6$Li+d channels are presented in Table \ref{tab_be8_li6}.
The $J^\pi=2^+$ channel cluster form factors are then shown in Fig. \ref{be8_li6_2p}.
We label the $J^\pi$ $^8$Be states as they are obtained in the current $6\hbar\Omega$ calculation.
The description of the excitation spectra of $^8$Be in the NCSM is generally very good
\cite{NCSM_8}. In Ref. \cite{NCSM_8} in addition to the $p$-shell states, the slowly
converging intruder $0^+ 0$, $2^+ 0$ and $4^+ 0$ states were found. Such states have 
complicated structure with wave functions dominated by higher than $0\hbar\Omega$ components.
Existence of such states was controversial \cite{FCB68,War,FZ98,FCBcom,FZrepcom,FCB00,AS88}. 
However, in the latest evaluation \cite{TUNL_A8}
a broad intruder $2^+$ state is included at about 9 MeV excitation energy. Such a state is
required by the R-matrix fits of nuclear reactions that involve $^8$Be
as the composite system. While in Ref. \cite{NCSM_8} the intruder states were investigated 
in the basis spaces up to $10\hbar\Omega$, in this paper we use the $6\hbar\Omega$ wave
functions to calculate the channel cluster form factors. In this space the intruder states
appear at a higher excitation energy and their importance is likely suppressed because of that.
In the present calculations, the $0^+ 0$ intruder state is the state $0^+_4$ and the $2^+ 0$ 
intruder state is the state $2^+_8$. Even in the current $6\hbar\Omega$ basis space, these
states have significant overlaps with the $^6$Li+d system. We note that the $^6$Li+d has a rather
high threshold of 22.28 MeV. Interestingly, we obtain a dominant overlap integral
in the $s=2$, $l=0$ channel for the $^8$Be $2^+$ excited state number seven, which is a $p$-shell
$0\hbar\Omega$-dominated state in our calculation with the excitation energy of 22.54 MeV.

\begin{table}
\begin{tabular}{c| c c| c c| c c| c c| c c}
\multicolumn{11} {c} {$\langle ^8$Be$| ^6$Li+d$\rangle$} \\
\hline
$J^\pi T$ & $(s,l)$ & $S$  & $(s,l)$ & $S$ & $(s,l)$ & $S$ & $(s,l)$ 
& $S$ & $(s,l)$ & $S$   \\
\hline
$0^+_1 0$  & $(0,0)$ & 1.051 & $(2,2)$ & 0.004 &  & & & & & \\
$0^+_3 0$  & $(0,0)$ & 0.111 & $(2,2)$ & 0.024 &  & & & & & \\
$0^+_4 0$  & $(0,0)$ & 0.194 & $(2,2)$ & 0.020 &  & & & & & \\
$0^+_5 0$  & $(0,0)$ & 0.393 & $(2,2)$ & 0.046 &  & & & & & \\
$2^+_1 0$  & $(2,0)$ & 0.004 & $(0,2)$ & 0.741 & $(1,2)$ & 0.009 & $(2,2)$ & 0.0005& $(2,4)$ & 0.003 \\
$2^+_3 0$  & $(2,0)$ & 0.137 & $(0,2)$ & 0.006 & $(1,2)$ & 0.010 & $(2,2)$ & 0.0007& $(2,4)$ & 0.0000\\
$2^+_4 0$  & $(2,0)$ & 0.290 & $(0,2)$ & 0.005 & $(1,2)$ & 0.471 & $(2,2)$ & 0.0012& $(2,4)$ & 0.0001\\
$2^+_7 0$  & $(2,0)$ & 0.442 & $(0,2)$ & 0.017 & $(1,2)$ & 0.136 & $(2,2)$ & 0.0005& $(2,4)$ & 0.0001\\
$2^+_8 0$  & $(2,0)$ & 0.015 & $(0,2)$ & 0.110 & $(1,2)$ & 0.002 & $(2,2)$ & 0.0003& $(2,4)$ & 0.001 \\
$2^+_9 0$  & $(2,0)$ & 0.0004& $(0,2)$ & 0.057 & $(1,2)$ & 0.021 & $(2,2)$ & 0.110 & $(2,4)$ & 0.0003\\
$2^+_{11} 0$&$(2,0)$ & 0.006 & $(0,2)$ & 0.035 & $(1,2)$ & 0.062 & $(2,2)$ & 0.136 & $(2,4)$ & 0.0003\\
$4^+_1 0$  & $(2,2)$ & 0.002 & $(0,4)$ & 0.037 & $(1,4)$ & 0.0000& $(2,4)$ & 0.0002& $(2,6)$ & 0.0005\\
$4^+_2 0$  & $(2,2)$ & 0.173 & $(0,4)$ & 0.001 & $(1,4)$ & 0.0000& $(2,4)$ & 0.0002& $(2,6)$ & 0.0000\\
$4^+_4 0$  & $(2,2)$ & 0.057 & $(0,4)$ & 0.006 & $(1,4)$ & 0.001 & $(2,4)$ & 0.0000& $(2,6)$ & 0.0000\\
$1^+_2 0$  & $(1,0)$ & 0.002 & $(1,2)$ & 0.025 & $(2,2)$ & 0.012 &  & & & \\
$1^+_3 0$  & $(1,0)$ & 0.008 & $(1,2)$ & 0.737 & $(2,2)$ & 0.003 &  & & & \\
$1^+_7 0$  & $(1,0)$ & 0.010 & $(1,2)$ & 0.009 & $(2,2)$ & 0.209 &  & & & \\
$3^+_1 0$  & $(1,2)$ & 0.563 & $(2,2)$ & 0.001 & $(1,4)$ & 0.0007& $(2,4)$ & 0.0003 & & \\
$3^+_3 0$  & $(1,2)$ & 0.0001& $(2,2)$ & 0.097 & $(1,4)$ & 0.0003& $(2,4)$ & 0.0000 & & \\
$3^+_5 0$  & $(1,2)$ & 0.013 & $(2,2)$ & 0.185 & $(1,4)$ & 0.0002& $(2,4)$ & 0.0003 & & \\
\end{tabular}
\caption{\label{tab_be8_li6} Spectroscopic factors for the $\langle ^8$Be$| ^6$Li+d$\rangle$
corresponding to the $^8$Be ground and excited states and the $^6$Li ground state. 
The CD-Bonn 2000 NN potential, the basis size of $N_{\rm max}=6$ and the HO frequency 
of $\hbar\Omega=13$ MeV were used. The $s$ and $l$ are the channel spin and the relative
angular momentum, respectively. The $0^+_4 0$ and $2^+_8 0$ are intruder states.
All other are $p$-shell states.
}
\end{table}

\begin{figure}
\vspace*{2cm}
\includegraphics[width=7.0in]{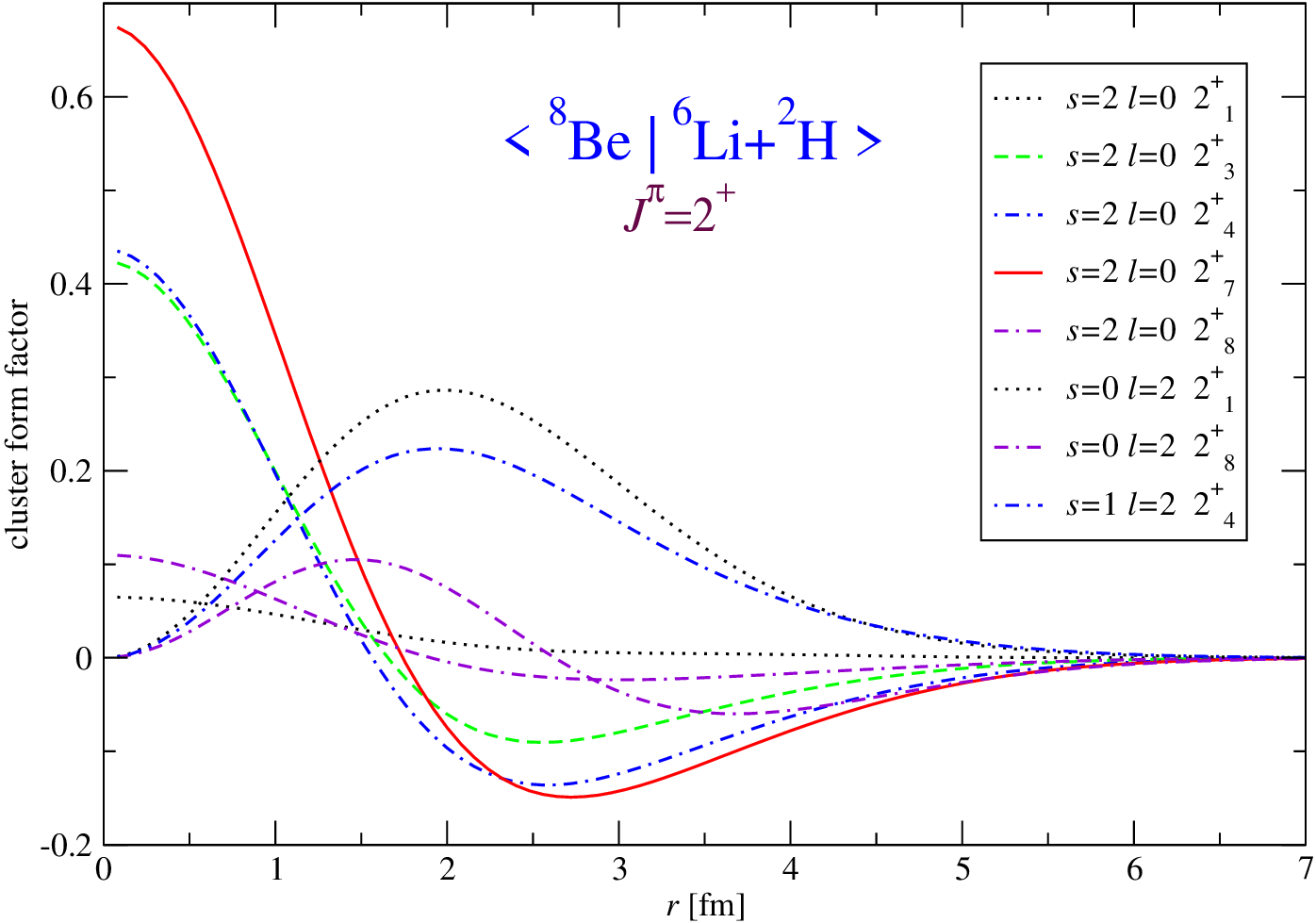}
\caption{\label{be8_li6_2p} (Color online)
Overlap integral of the $^8$Be $2^+$ states with the
$^6$Li+d as a function of separation between the $^6$Li and the deuteron.
The CD-Bonn 2000 NN potential, the basis size of $N_{\rm max}=6$ and the HO frequency 
of $\hbar\Omega=13$ MeV were used. The $2^+_8 0$ is an intruder state.
All other are $p$-shell states.
}
\end{figure}

We note a technical issue affecting the $\langle ^{8}$Be$| ^{6}$Li+d$\rangle$ overlap integral 
calculations using Eq. (\ref{two-nucleon}). When an m-scheme calculation is performed
with a fixed $M$, it is in general necessary to use $M>0$ for $J>0$ channels
in order to generate all needed reduced matrix elements of the $a^\dagger a^\dagger$ 
operators.

\subsection{$\langle ^{8}$Be$| ^{7}$Li+p$\rangle$}\label{be8_li7}

Our spectroscopic factors for the $\langle ^{8}$Be$| ^{7}$Li+p$\rangle$ channels
are presented in Table \ref{tab_be8_li7}. The overlap integrals for the $J^\pi=1^+$ channels
are shown in Fig. \ref{be8_li7_1p}. 
It should be noted that the expressions for cluster form factors and spectroscopic
factors presented in Sect. \ref{sec_calc} employ isospin formalism. In order to distinguish, 
e.g. a proton or a neutron projectile, the Eq. (\ref{single-nucleon}) must be multiplied
by the isospin Clebsch-Gordan coefficient $(T_1 M_{T_1} \frac{1}{2} M_{T_2} | T M_T)$
with $M_{T_2}=+\frac{1}{2}$ ($-\frac{1}{2}$) for proton (neutron) \cite{Brussaard}. 
This coefficient is 1 for all the overlaps studied in this paper except 
the $\langle ^{8}$Be$| ^{7}$Li+p$\rangle$ overlap for which it is equal 
to $-\frac{1}{\sqrt{2}},\frac{1}{\sqrt{2}}$ for $T=0,1$ $^8$Be states, respectively. 
We find large overlap integrals for the $1^+_1$ 
and the $1^+_2$ states as well as large spectroscopic factors for the $3^+_1$ state
consistent with the resonances in the $^7$Li+p cross section \cite{AS88,TUNL_A8}.
We also note very large spectroscopic factors and cluster overlap integrals
for the $1^+_4$ state which is the second $T=1$ $1^+$ state in our calculation appearing at
the excitation energy of 20.37 MeV. Such a state is not included in the current evaluations
\cite{AS88,TUNL_A8}. It is, however, needed in the R-matrix analysis \cite{Page}. 
Comparing our spectroscopic factors to the Cohen-Kurath calculations \cite{CK_spec},
we have a reasonable agreement for the lowest states
(after correcting for the above discussed factor of 2 
due to the isospin Clebsch-Gordan coefficient), with increasing differences for the
higher lying states. Our results are also influenced by the isospin mixing of the $2^+_2$, $2^+_3$
and $1^+_1$, $1^+_2$ states. 

\begin{figure}
\vspace*{2cm}
\includegraphics[width=7.0in]{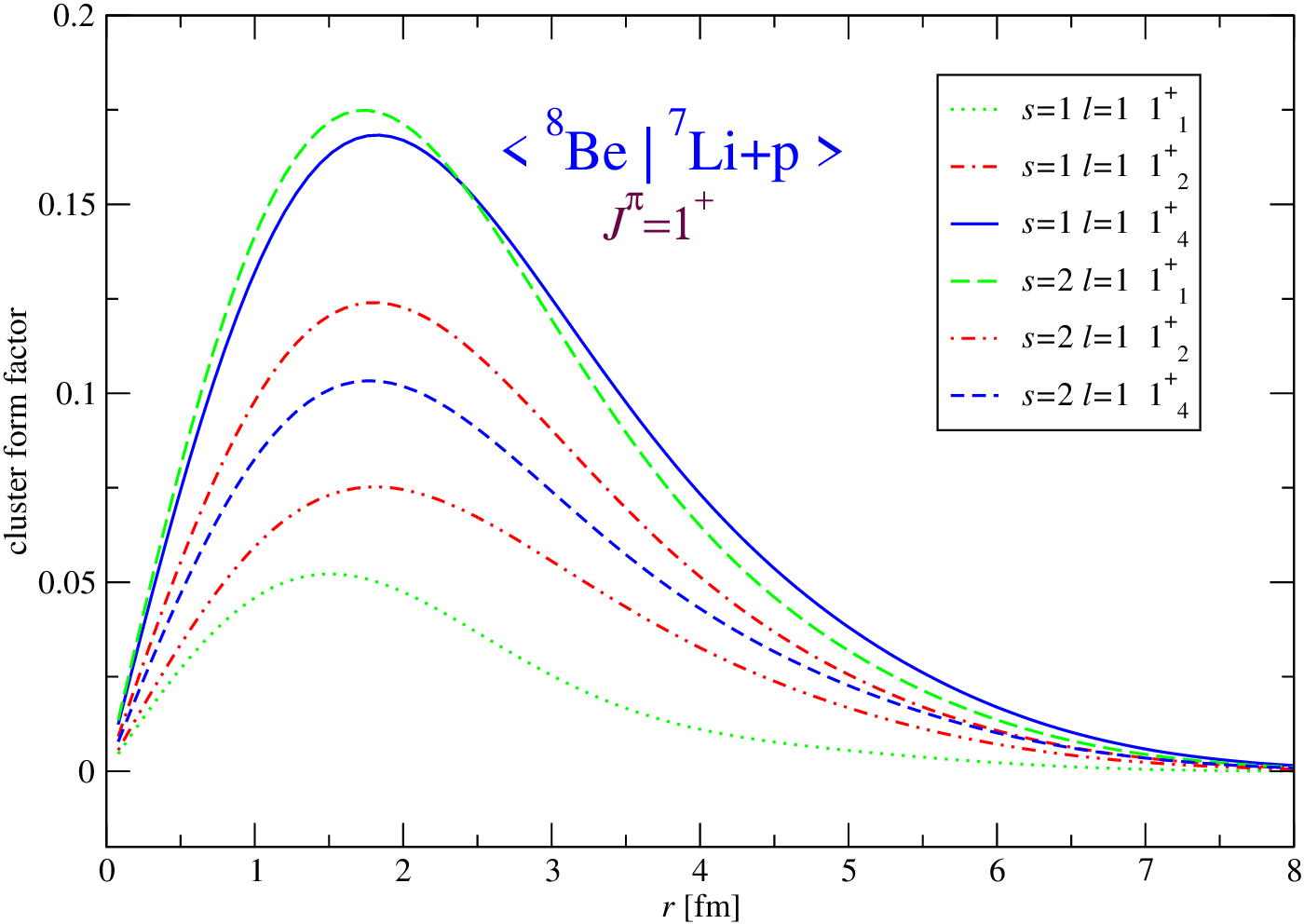}
\caption{\label{be8_li7_1p} (Color online)
Overlap integral of the $^8$Be $1^+$ states with the
$^7$Li+p as a function of separation between the $^7$Li and the proton.
The CD-Bonn 2000 NN potential, the basis size of $N_{\rm max}=6$ and the HO frequency 
of $\hbar\Omega=13$ MeV were used.
}
\end{figure}

\begin{table}
\begin{tabular}{c| c c| c c| c c| c c}
\multicolumn{9} {c} {$\langle ^8$Be$| ^7$Li+p$\rangle$} \\
\hline
$J^\pi T$ & $(s,l)$ & $S$  & $(s,l)$ & $S$ & $(s,l)$ & $S$ & $(s,l)$ & $S$ \\
\hline
$0^+_1 0$  & $(1,1)$ & 1.520 &         &       &  & & & \\
$0^+_2 1$  & $(1,1)$ & 0.192 &         &       &  & & & \\
$0^+_3 0$  & $(1,1)$ & 0.144 &         &       &  & & & \\
$0^+_4 0$  & $(1,1)$ & 0.212 &         &       &  & & & \\
$0^+_5 0$  & $(1,1)$ & 0.006 &         &       &  & & & \\
$2^+_1 0$  & $(1,1)$ & 0.913 & $(2,1)$ & 0.007 & $(1,3)$ & 0.018 & $(2,3)$ & 0.0000\\
$2^+_2 1$  & $(1,1)$ & 0.157 & $(2,1)$ & 0.629 & $(1,3)$ & 0.0000& $(2,3)$ & 0.001 \\
$2^+_3 0$  & $(1,1)$ & 0.018 & $(2,1)$ & 0.194 & $(1,3)$ & 0.001 & $(2,3)$ & 0.0000\\
$2^+_4 0$  & $(1,1)$ & 0.050 & $(2,1)$ & 0.060 & $(1,3)$ & 0.0000& $(2,3)$ & 0.0005\\
$2^+_5 1$  & $(1,1)$ & 0.059 & $(2,1)$ & 0.164 & $(1,3)$ & 0.0001& $(2,3)$ & 0.0025\\
$2^+_6 1$  & $(1,1)$ & 0.102 & $(2,1)$ & 0.015 & $(1,3)$ & 0.001 & $(2,3)$ & 0.0005\\
$2^+_7 0$  & $(1,1)$ & 0.004 & $(2,1)$ & 0.062 & $(1,3)$ & 0.002 & $(2,3)$ & 0.002 \\
$2^+_8 0$  & $(1,1)$ & 0.049 & $(2,1)$ & 0.001 & $(1,3)$ & 0.005 & $(2,3)$ & 0.0003\\
$4^+_1 0$  & $(1,3)$ & 0.023 & $(2,3)$ & 0.0000& $(1,5)$ & 0.0001& $(2,5)$ & 0.0000\\
$1^+_1 1$  & $(1,1)$ & 0.020 & $(2,1)$ & 0.367 & $(2,3)$ & 0.0001&  & \\
$1^+_2 0$  & $(1,1)$ & 0.207 & $(2,1)$ & 0.080 & $(2,3)$ & 0.002 &  & \\
$1^+_3 0$  & $(1,1)$ & 0.005 & $(2,1)$ & 0.002 & $(2,3)$ & 0.006 &  & \\
$1^+_4 1$  & $(1,1)$ & 0.404 & $(2,1)$ & 0.145 & $(2,3)$ & 0.0000&  & \\
$1^+_5 1$  & $(1,1)$ & 0.0005& $(2,1)$ & 0.0085& $(2,3)$ & 0.0015&  & \\
$3^+_1 0$  & $(2,1)$ & 0.322 & $(1,3)$ & 0.0015& $(2,3)$ & 0.002 & $(2,5)$ & 0.0001 \\
$3^+_2 1$  & $(2,1)$ & 0.090 & $(1,3)$ & 0.0000& $(2,3)$ & 0.001 & $(2,5)$ & 0.0002 \\
$3^+_3 0$  & $(2,1)$ & 0.002 & $(1,3)$ & 0.0005& $(2,3)$ & 0.004 & $(2,5)$ & 0.0000 \\
$3^+_4 1$  & $(2,1)$ & 0.013 & $(1,3)$ & 0.0000& $(2,3)$ & 0.0000& $(2,5)$ & 0.0000 \\
\end{tabular}
\caption{\label{tab_be8_li7} Spectroscopic factors for the $\langle ^8$Be$| ^7$Li+p$\rangle$
corresponding to the $^8$Be ground and excited states and the $^7$Li ground state. 
The CD-Bonn 2000 NN potential, the basis size of $N_{\rm max}=6$ and the HO frequency 
of $\hbar\Omega=13$ MeV were used. The $s$ and $l$ are the channel spin and the relative
angular momentum, respectively. The $0^+_4 0$ and $2^+_8 0$ are intruder states.
All other are $p$-shell states.
}
\end{table}

\subsection{$\langle ^{9}$Li$| ^{8}$Li+n$\rangle$}\label{li9_li8}

The experimental information on $^9$Li is rather limited \cite{AS88,TUNL_A9}.
New experiments are under way or planned, however, to explore this nucleus.
One of such experiments is the inverse-kinematic d($^8$Li,p) scattering \cite{Wuosmaa}.
It is therefore useful to perform theoretical calculations of the 
$\langle ^{9}$Li$| ^{8}$Li+n$\rangle$ spectroscopic factors. Our results obtained for both
the negative and the positive parity states of $^9$Li are summarized in Table \ref{tab_li9_li8}.
In the present calculation we employed the AV8$^\prime$+TM$^\prime$(99) two- plus 
three-nucleon interaction in the NCSM calculations performed in an approach described
in Ref. \cite{NCSM_v3b}. The Tucson-Melbourne (TM) three-nucleon interaction
was introduced in Ref. \cite{TM} with the particular version we are using, the TM$^\prime(99)$, 
described in Ref. \cite{TMprime99}. The Argonne V8$^\prime$ NN potential is a slightly 
simplified version of the high-quality AV18 interaction \cite{GFMC}. Our current calculations
are limited to the $4\hbar\Omega$ and $5\hbar\Omega$ basis spaces for the negative 
and the positive parity states, respectively, due to the complexity of the calculation
with a genuine three-nucleon interaction. In general, it is accepted that a three-nucleon interaction
is needed in addition to the high-quality NN potentials to explain the few-nucleon system 
binding energies and to improve description of some three-nucleon scattering observables.
Recently, it has been shown that the three-nucleon interaction is also needed for a correct
description of low-lying excitation spectra of $p$-shell nuclei \cite{GFMC,NCSM_v3b}.

In Fig. \ref{li9_li8_5m}, we show the channel cluster form factor for the lowest two
$\frac{5}{2}^-$ states. The $\frac{5}{2}^-_1$ state with a large spectroscopic factor
is a candidate for the 4.296 MeV $^9$Li state lying just above the 4.063 MeV
$^6$Li+n threshold.  

\begin{figure}[p]
\vspace*{2cm}
\includegraphics[width=7.0in]{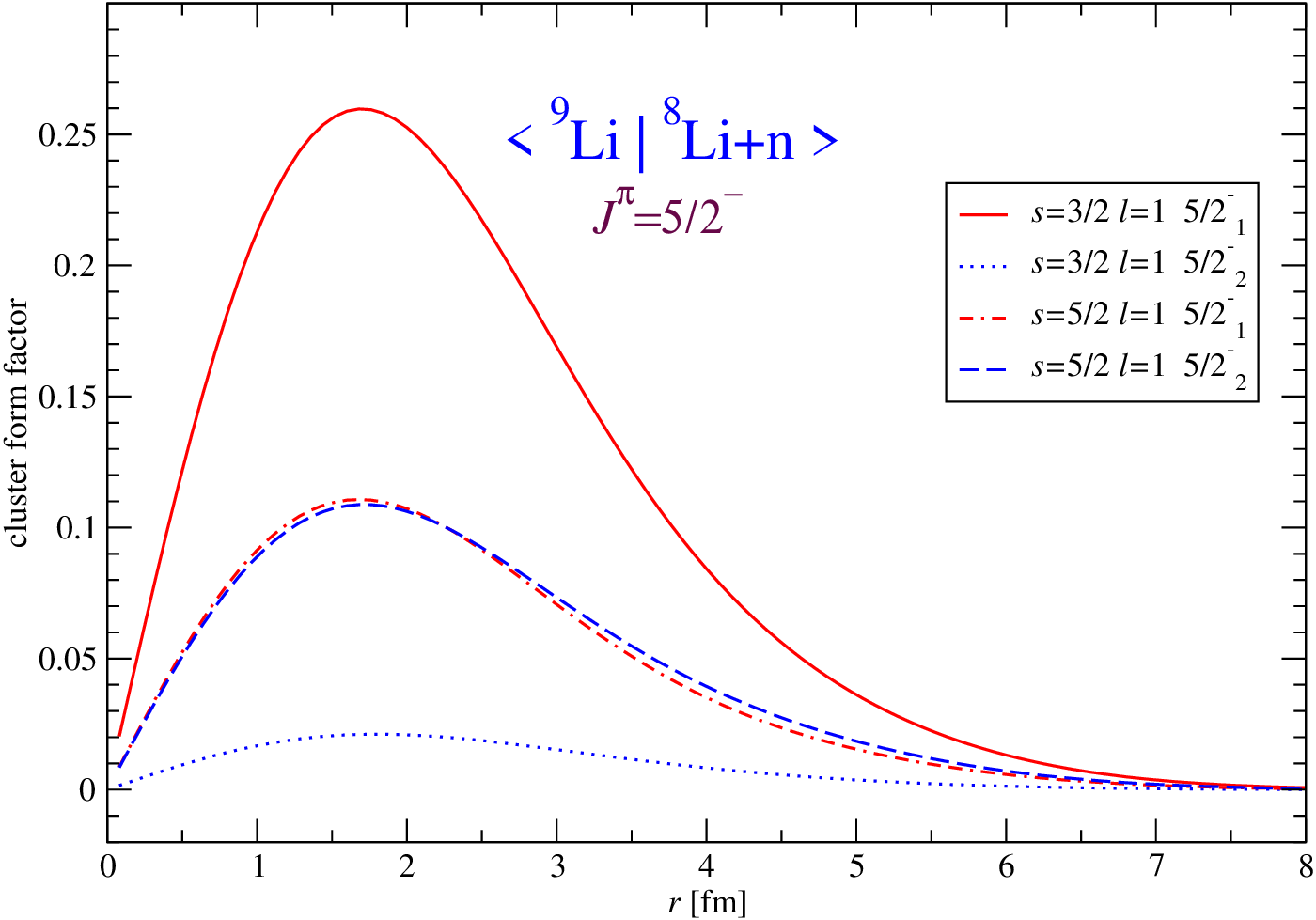}
\caption{\label{li9_li8_5m} (Color online)
Overlap integral of the $^9$Li two lowest $\frac{5}{2}^-$ states with the
$^8$Li+n as a function of separation between the $^8$Li and the neutron.
The AV8'+TM'(99) two- plus three-nucleon interaction, the basis size of $N_{\rm max}=4$ 
and the HO frequency of $\hbar\Omega=14$ MeV were used.
}
\end{figure}

\begin{table}
\begin{tabular}{c| c c| c c| c c}
\multicolumn{7} {c} {$\langle ^9$Li$| ^8$Li+n$\rangle$} \\
\hline
$J^\pi T$ & $(s,l)$ & $S$  & $(s,l)$ & $S$ & $(s,l)$ & $S$   \\
\hline
$\frac{3}{2}^-_1 \frac{3}{2}$ & $(\frac{3}{2},1)$ & 0.628 & $(\frac{5}{2},1)$ & 0.426 & & \\
$\frac{1}{2}^-_1 \frac{3}{2}$ & $(\frac{3}{2},1)$ & 0.519 & $(\frac{5}{2},3)$ & 0.005 & & \\
$\frac{5}{2}^-_1 \frac{3}{2}$ & $(\frac{3}{2},1)$ & 0.711 & $(\frac{5}{2},1)$ & 0.126 & & \\
$\frac{3}{2}^-_2 \frac{3}{2}$ & $(\frac{3}{2},1)$ & 0.030 & $(\frac{5}{2},1)$ & 0.180 & & \\
$\frac{7}{2}^-_1 \frac{3}{2}$ & $(\frac{5}{2},1)$ & 0.0006& $(\frac{3}{2},3)$ & 0.0002& 
$(\frac{5}{2},3)$& 0.001 \\
$\frac{3}{2}^-_3 \frac{3}{2}$ & $(\frac{3}{2},1)$ & 0.048 & $(\frac{5}{2},1)$ & 0.703 & & \\
$\frac{5}{2}^-_2 \frac{3}{2}$ & $(\frac{3}{2},1)$ & 0.006 & $(\frac{5}{2},3)$ & 0.137 & & \\
$\frac{1}{2}^-_2 \frac{3}{2}$ & $(\frac{3}{2},1)$ & 0.135 &   & & & \\
$\frac{5}{2}^+_1 \frac{3}{2}$ & $(\frac{5}{2},0)$ & 0.790 & $(\frac{3}{2},2)$ & 0.122 
& $(\frac{5}{2},2)$ & 0.038  \\  
$\frac{3}{2}^+_1 \frac{3}{2}$ & $(\frac{3}{2},0)$ & 0.686 & $(\frac{3}{2},2)$ & 0.008 
& $(\frac{5}{2},2)$ & 0.017  \\
$\frac{1}{2}^+_1 \frac{3}{2}$ & $(\frac{3}{2},2)$ & 0.025 & $(\frac{5}{2},2)$ & 0.011 & & \\
$\frac{7}{2}^+_1 \frac{3}{2}$ & $(\frac{3}{2},2)$ & 0.164 & $(\frac{5}{2},2)$ & 0.315 
& $(\frac{5}{2},4)$ & 0.001  \\
$\frac{3}{2}^+_2 \frac{3}{2}$ & $(\frac{3}{2},0)$ & 0.127 & $(\frac{3}{2},2)$ & 0.183 
& $(\frac{5}{2},2)$ & 0.005  \\
$\frac{5}{2}^+_2 \frac{3}{2}$ & $(\frac{5}{2},0)$ & 0.133 & $(\frac{3}{2},2)$ & 0.107 
& $(\frac{5}{2},2)$ & 0.008  \\
$\frac{3}{2}^+_3 \frac{3}{2}$ & $(\frac{3}{2},0)$ & 0.001 & $(\frac{3}{2},2)$ & 0.130 
& $(\frac{5}{2},2)$ & 0.009  \\  
$\frac{9}{2}^+_1 \frac{3}{2}$ & $(\frac{5}{2},2)$ & 0.713 & $(\frac{5}{2},4)$ & 0.002 & & \\
\end{tabular}
\caption{\label{tab_li9_li8} Spectroscopic factors for the $\langle ^9$Li$| ^8$Li+n$\rangle$
corresponding to the $^9$Li ground and excited states and the $^8$Li $2^+ 1$ ground state. 
The AV8$^\prime$+TM$^\prime$(99) two- plus three-body interaction, 
the basis size of $N_{\rm max}=4,5$ and the HO frequency 
of $\hbar\Omega=14$ MeV were used. The $s$ and $l$ are the channel spin and the relative
angular momentum, respectively. Only channels with $S\geq 0.001$ are shown.
}
\end{table}

\subsection{$\langle ^{13}$C$| ^{12}$C+n$\rangle$}\label{c13_c12}

Apart from an increase of the binding energy, the genuine three-nucleon interaction
also causes an increase of the spin-orbit splitting. This is demonstrated not only in 
different level spacing and sometimes in a different level ordering in calculations
with the three-nucleon interaction, e.g. in $^{10}$B, 
but also in the spectroscopic factors and the overlap 
integrals as we show in this subsection for $\langle ^{13}$C$| ^{12}$C+n$\rangle$ system. 
Using the wave functions obtained in Ref. \cite{NCSM_v3b}, we compare 
in Fig. \ref{c13_c12_comp} and  Table \ref{tab_c13_c12} the
$\frac{1}{2}^-$ and $\frac{3}{2}^-$ channel cluster form factors and the spectroscopic
factors, respectively, obtained in calculations with and without the TM$^\prime$(99) 
three-nucleon interaction. The AV8$^\prime$ NN potential is used for the two-nucleon 
interaction. We can see that the $\frac{1}{2}^-$ channel cluster form factor and the 
spectroscopic factor increase when the three-nucleon interaction is included, while 
at the same time the $\frac{3}{2}^-$ factors decrease. This can be understood as 
an increase of the spin-orbit splitting of the $0p_{\frac{3}{2}}$-$0p_{\frac{1}{2}}$
levels due to the three-nucleon interaction which results in a purer $(0p_{\frac{3}{2}})^8$
$^{12}$C ground state and a purer $(0p_{\frac{3}{2}})^8, (0p_{\frac{1}{2}})^1$ $^{13}$C 
ground state.

\begin{figure}[p]
\vspace*{2cm}
\includegraphics[width=7.0in]{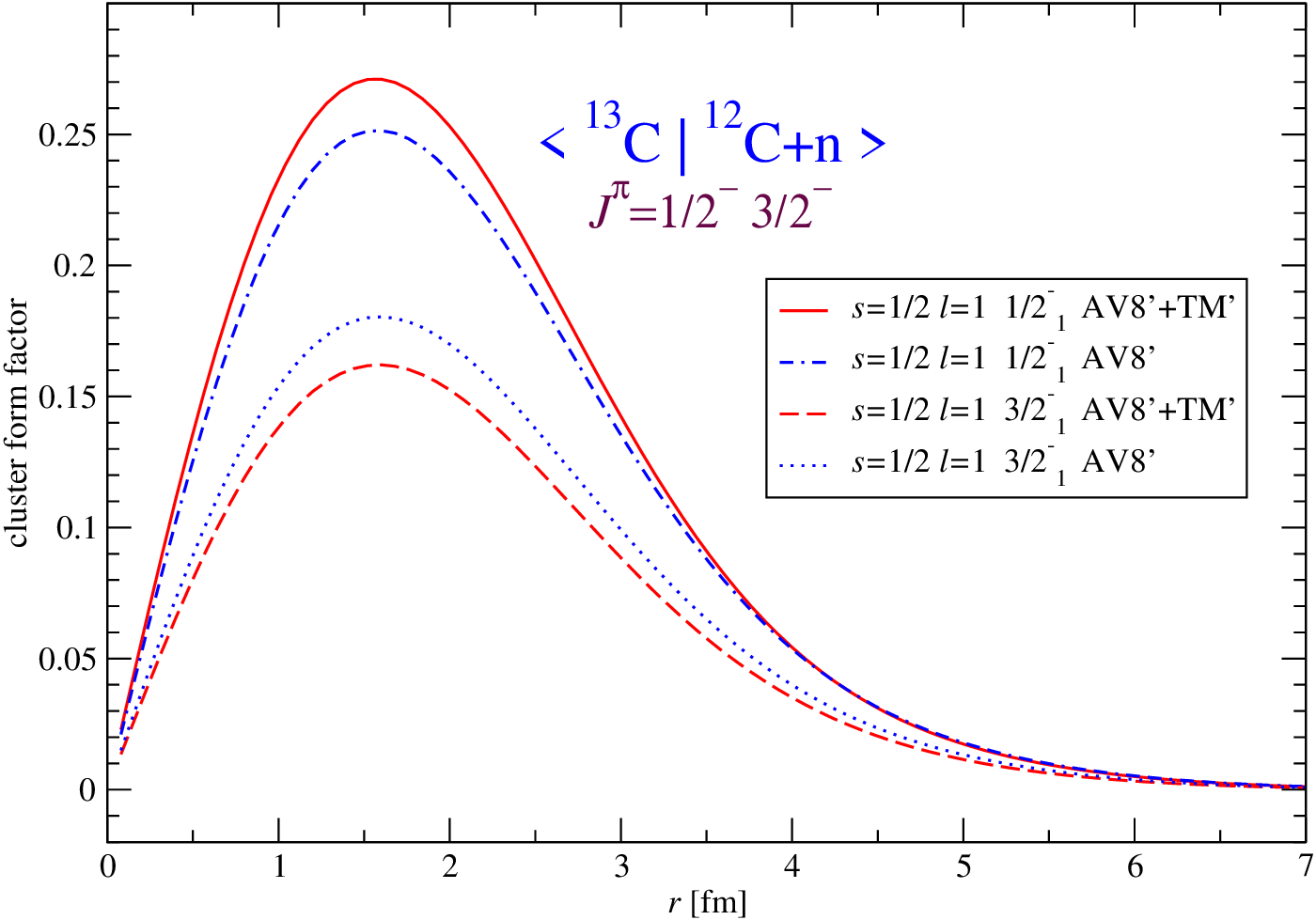}
\caption{\label{c13_c12_comp} (Color online)
Overlap integral of the $^{13}$C $\frac{1}{2}^-$ ground state and the $\frac{3}{2}^-$ first 
excited state with the $^{12}$C+n as a function of separation between the $^{12}$C and the neutron.
Results obtained using the AV8' NN potential and the AV8'+TM'(99) two- plus three-nucleon interaction
are compared. The basis size of $N_{\rm max}=4$ and the HO frequency of $\hbar\Omega=15$ MeV were used.
}
\end{figure}

\begin{table}
\begin{tabular}{c| c c | c c}
\multicolumn{5} {c} {$\langle ^{13}$C$| ^{12}$C+n$\rangle$} \\
\hline
           & \multicolumn{2} {c} {AV8$^\prime$+TM$^\prime$(99)}  &
\multicolumn{2} {c} {AV8$^\prime$} \\
$J^\pi T$ & $(s,l)$ & $S$ & $(s,l)$ & $S$  \\
\hline
$\frac{1}{2}^-_1 \frac{1}{2}$ & $(\frac{1}{2},1)$ & 0.549 & $(\frac{1}{2},1)$ & 0.489 \\
$\frac{3}{2}^-_1 \frac{1}{2}$ & $(\frac{1}{2},1)$ & 0.206 & $(\frac{1}{2},1)$ & 0.258 \\
$\frac{1}{2}^-_2 \frac{1}{2}$ & $(\frac{1}{2},1)$ & 0.015 & $(\frac{1}{2},1)$ & 0.002 \\
$\frac{3}{2}^-_2 \frac{1}{2}$ & $(\frac{1}{2},1)$ & 0.001 & $(\frac{1}{2},1)$ & 0.0005 \\
$\frac{1}{2}^-_3 \frac{1}{2}$ & $(\frac{1}{2},1)$ & 0.034 & $(\frac{1}{2},1)$ & 0.016 \\
$\frac{3}{2}^-_3 \frac{1}{2}$ & $(\frac{1}{2},1)$ & 0.008 & $(\frac{1}{2},1)$ & 0.004 \\
\end{tabular}
\caption{\label{tab_c13_c12} Spectroscopic factors for the $\langle ^{13}$C$| ^{12}$C+n$\rangle$
corresponding to the $^{13}$C ground and excited states and the $^{12}$C ground state. 
Results obtained using the AV8$^\prime$+TM$^\prime$(99) two- plus three-body interaction
and the AV8$^\prime$ NN interaction are compared.
The basis size of $N_{\rm max}=4$ and the HO frequency of $\hbar\Omega=15$ MeV were used.
The $s$ and $l$ are the channel spin and the relative angular momentum, respectively. 
All the presented $^{13}$C states are $p$-shell states.
}
\end{table}

Comparing to the phenomenological Cohen-Kurath spectroscopic factors \cite{CK_spec},
a better agreement is achieved in the more realistic calculation with the three-nucleon
interaction.

\section{Conclusions}\label{sec_concl}

We derived expressions for calculations of channel cluster form factors and spectroscopic 
factors from the {\it ab initio} no-core shell model wave functions. We considered 
the most practical case, with the composite system and the target nucleus described 
in the Slater determinant harmonic oscillator basis while the projectile eigenstate 
expanded in the Jacobi coordinate HO basis. The spurious center of mass components
present in the SD bases were removed exactly. The calculated cluster form factors are then 
translationally invariant. The algebraic expressions for the channel cluster form factors were 
derived for up to four-nucleon projectiles. We numerically tested these expressions for systems
consisting of up to a three-nucleon projectile. Several numerical tests were performed that 
involved interchanging the role of the target and the projectile as well as performance of two 
independent calculations one of which employed only the Jacobi-coordinate wave functions
for all nuclei involved while the other used the SD basis wave functions for the composite
system and the target. Identical results were obtained in both cases.   

As examples of application, we presented results 
for $\langle ^5$He$| ^4$He+n$\rangle$, $\langle ^5$He$| ^3$H+d$\rangle$, 
$\langle ^6$Li$| ^4$He+d$\rangle$,
$\langle ^6$Be$| ^3$He$+^3$He$\rangle$, $\langle ^7$Li$| ^4$He$+^3$H$\rangle$, 
$\langle ^7$Li$| ^6$Li+n$\rangle$,
$\langle ^8$Be$| ^6$Li+d$\rangle$, $\langle ^8$Be$| ^7$Li+p$\rangle$, 
$\langle ^9$Li$| ^8$Li+n$\rangle$ and
$\langle ^{13}$C$| ^{12}$C+n$\rangle$ systems, with all the nuclei described 
by multi-$\hbar\Omega$ 
NCSM wave functions. The calculations involve no fitting. Apart from the basis size,
the only parameter appearing in the NCSM is the HO frequency, which is typically fixed
so that the binding energy is the least dependent on the HO frequency. In the current
application, this is hard to achieve in the cases that involve both the $0s$- and the 
$0p$-shell nuclei, as we require the same HO frequency for all nuclei.
Therefore, we studied the dependence on the basis size and on the HO frequency 
in most investigated cases. It is very encouraging that our results ere rather stable 
and robust. Additionally, we found a qualitative agreement with experiment for, e.g.,
$\langle ^5$He$| ^4$He+n$\rangle$, $\langle ^5$He$| ^3$H+d$\rangle$, 
$\langle ^7$Li$| ^4$He$+^3$H$\rangle$, $\langle ^7$Li$| ^6$Li+n$\rangle$, 
$\langle ^8$Be$| ^7$Li+p$\rangle$, 
in the sense that large channel cluster form factors correspond to resonances in cross 
sections. This confirms that the multi-$\hbar\Omega$ NCSM wave functions provide
a realistic description of light nuclei, in particular for the low-lying $p$-shell
states. 

As a further development, apart from performing calculations for systems with 
a four-nucleon projectile, our goal is to utilize the channel cluster form factors
as a first step to describe low energy reactions on light nuclei. The presently calculated
channel cluster form factors were obtained using the model space wave functions.
As the next step, we need to take into account the influence of the complementary space
and calculate effective, or renormalized, channel cluster form factors. It is expected
that this will improve the cluster form factors at intermediate distances and make them
more suitable for matching to the correct asymptotic cluster wave functions.
Hopefully, it will be possible to develop a microscopic nuclear reaction approach
similar to the RGM \cite{RGM} starting, however, 
from realistic {\it ab initio} wave functions.

\section{Acknowledgments}

This work was performed under the auspices of
the U. S. Department of Energy by the University of California,
Lawrence Livermore National Laboratory under contract
No. W-7405-Eng-48. Support from the LDRD contract No. 04-ERD-058 
as well as partial support from the DOE grant SCW0498 is acknowledged.

\appendix

\section{$12-j$ symbol definition}\label{appA}

The $12-j$ symbol of the first kind \cite{Varshalovich} is defined by
\begin{equation}\label{12j}
\left\{ \begin{array}{cccc} 
     a & d & e & h \\
     p & q & r & s \\
     b & c & f & g
\end{array}\right\} = \sum_X (-1)^{a+b+c+d+e+f+g+h+p+q+r+s-X} \hat{X}^2
\left\{ \begin{array}{ccc} 
  a & b & X \\
  c & d & p   
\end{array}\right\}
\left\{ \begin{array}{ccc} 
  c & d & X \\
  e & f & q   
\end{array}\right\}
\left\{ \begin{array}{ccc} 
  e & f & X \\
  g & h & r   
\end{array}\right\}
\left\{ \begin{array}{ccc} 
  g & h & X \\
  b & a & s   
\end{array}\right\}
\; .
\end{equation}

\section{Four-nucleon projectile}\label{appB}

The channel cluster form factor expression for the case of a four-nucleon projectile is 
\begin{eqnarray}\label{four-nucleon}
\langle A \lambda JT |&{\cal A}& \Phi_{\alpha I_1 T_1,\beta I_2 T_2; s l}^{(A-4,4) JT};
\delta_{\eta_{A-4}}\rangle
= \sum_n R_{nl}(\eta_{A-4})\frac{1}{\sqrt{24}} 
\frac{1}{\langle nl00l|00nll\rangle_{\frac{4}{A-4}}} \frac{1}{\hat{J}\hat{T}} 
\nonumber \\
&\times&
\sum \langle (((n_2 l_2 s_2 j_2 t_2;{\cal N}_2 {\cal L}_2 {\cal J}_2 \textstyle{\frac{1}{2}}) J_3 T_3)
{\cal N}_3 {\cal L}_3 {\cal J}_3 \textstyle{\frac{1}{2}}) I_2 T_2|a=4 \beta I_2 T_2 \rangle
\nonumber \\
&\times& 
\hat{s}\hat{I}\hat{s}_2 \hat{j}_2 \hat{I}_2 \hat{I}_3 \hat{\cal J}_2 \hat{\cal J}_3 \hat{J}_3
\hat{j}_a\hat{j}_b \hat{j}_c \hat{j}_d \hat{I}_{ab} \hat{\lambda}^2 \hat{\kappa}^2 \hat{L}_{ab}^2 
(-1)^{I_1+l+J+l_2+t_2+{\cal J}_2+{\cal J}_3+l_c+l_d+I_{ab}+I}
\left\{ \begin{array}{ccc} I_1 & I_2 & s \\
  l  & J & I   
\end{array}\right\}
\nonumber \\
&\times&
\left\{ \begin{array}{ccc} L_2 & L_{ab} & l_2 \\
  s_2  & j_2 & I_{ab}   
\end{array}\right\}
\left\{ \begin{array}{ccc} 
     l_a & l_b & L_{ab} \\
  \textstyle{\frac{1}{2}} & \textstyle{\frac{1}{2}} & s_2 \\
     j_a & j_b & I_{ab}
\end{array}\right\}
\left\{ \begin{array}{cccc} 
     L_3 & \lambda & L_2 & j_2 \\
   {\cal L}_2  & l_c & I_{ab} & J_3 \\
   {\cal J}_2 & \textstyle{\frac{1}{2}} & j_c & I_3
\end{array}\right\}
\left\{ \begin{array}{cccc} 
     l & \kappa & L_3 & J_3 \\
   {\cal L}_3  & l_d & I_3 & I_2 \\
   {\cal J}_3 & \textstyle{\frac{1}{2}} & j_d & I
\end{array}\right\}
\nonumber \\
&\times&
\langle n_a l_a n_b l_b L_{ab}|N_2 L_2 n_2 l_2 L_{ab}\rangle_{1}
\langle n_c l_c N_2 L_2 \lambda|N_3 L_3 {\cal N}_2 {\cal L}_2 \lambda \rangle_{\textstyle{\frac{1}{2}}}
\langle n_d l_d N_3 L_3 \kappa|n l {\cal N}_3 {\cal L}_3 \kappa \rangle_{\textstyle{\frac{1}{3}}}
\nonumber \\
&\times&
\; _{\rm SD}\langle A\lambda JT|||(((a^\dagger_{n_a l_a j_a} a^\dagger_{n_b l_b j_b})^{(I_{ab} t_2)}
a^\dagger_{n_c l_c j_c})^{(I_3 T_3)} a^\dagger_{n_d l_d j_d})^{(I T_2)}
|||A-4\alpha I_1 T_1\rangle_{\rm SD} \;
\; .
\end{eqnarray}
Similarly as the three-nucleon eigenstates in Eqs. (\ref{three-nucleon},\ref{cfp_3}), 
the four-nucleon eigenstates are expanded in a basis with lower degree of antisymmetry
using the coefficients of fractional parentage \cite{Jacobi_NCSM}
\begin{eqnarray}\label{cfp_4}
\langle (((n_2 l_2 s_2 j_2 t_2;{\cal N}_2 {\cal L}_2 {\cal J}_2 \textstyle{\frac{1}{2}}) J_3 T_3)
{\cal N}_3 {\cal L}_3 {\cal J}_3 \textstyle{\frac{1}{2}}) I_2 T_2|a=4 \beta I_2 T_2 \rangle
&=& \sum 
\langle n_2 l_2 s_2 j_2 t_2;{\cal N}_2 {\cal L}_2 {\cal J}_2 \textstyle{\frac{1}{2}} 
|| N_{x} i_{x} J_3 T_3 \rangle
\nonumber \\
&\times&
\langle N_{x} i_{x} J_3 T_3; {\cal N}_3 {\cal L}_3 {\cal J}_3 \textstyle{\frac{1}{2}})
|| N i I_2 T_2 \rangle
\nonumber \\
&\times&
\langle N i I_2 T_2 | a=4 \beta I_2 T_2 \rangle 
\; ,
\end{eqnarray}
with $N_{x}=2{\cal N}_2+{\cal L}_2 + 2n_2 + l_2$ and $N=N_{x}+2{\cal N}_3+{\cal L}_3$
the total number of HO excitations 
for the three and four nucleons and $i_{x},i$ the additional quantum numbers that characterize 
the three- and four-nucleon antisymmetrized basis states, respectively.
In the case of the $^4$He projectile, $I_2=T_2=0$ and $(-1)^{l_2+{\cal L}_2+{\cal L}_3}=1$.
In Eq. (\ref{four-nucleon}), in addition to the HO bracket (\ref{cm_ho_br}) due 
to the CM correction, three general HO brackets appear that correspond
to particles with mass ratios $1,\frac{1}{2}$ and $\frac{1}{3}$.
These are due to the sequence of three transformations of the HO wave functions
$\varphi_{nlm}(\vec{R}_{\rm CM}^{a=3}) \varphi_{n_2 l_2 m_2}(\vec{\vartheta}_{A-1}) 
\varphi_{{\cal N}_2 {\cal L}_2 {\cal M}_2}(\vec{\vartheta}_{A-2})
\varphi_{{\cal N}_3 {\cal L}_3 {\cal M}_3}(\vec{\vartheta}_{A-3})$ 
to the single-particle HO wave functions 
$\varphi_{n_a l_a m_a}(\vec{r}_A) \varphi_{n_b l_b m_b}(\vec{r}_{A-1}) 
\varphi_{n_c l_c m_c}(\vec{r}_{A-2}) \varphi_{n_d l_d m_d}(\vec{r}_{A-3})$.
With the help of Eq. (\ref{four-nucleon}), one can study the alpha-cluster structure
of the $p$-shell nuclei states. In the past, this has been typically investigated 
using the cluster models, see e.g. Ref. \cite{Arai04}.

\end{document}